\documentclass[11pt]{article}
\usepackage{amsmath,amsthm,latexsym,amssymb,amsfonts,epsfig}

\makeatletter
\@addtoreset{equation}{section}
\makeatother

\def\be{\begin{equation}}
\def\ee{\end{equation}}
\def\ba{\begin{eqnarray}}
\def\ea{\end{eqnarray}}

\newcommand\nn{\nonumber}
\newcommand\q{\quad}
\def\Nl{{\mathchoice
{\setbox0=\hbox{$\displaystyle\rm N$}\hbox{\hbox to0pt
{\kern0.4\wd0\vrule height0.9\ht0\hss}\box0}}
{\setbox0=\hbox{$\textstyle\rm N$}\hbox{\hbox to0pt
{\kern0.4\wd0\vrule height0.9\ht0\hss}\box0}}
{\setbox0=\hbox{$\scriptstyle\rm N$}\hbox{\hbox to0pt
{\kern0.4\wd0\vrule height0.9\ht0\hss}\box0}}
{\setbox0=\hbox{$\scriptscriptstyle\rm N$}\hbox{\hbox to0pt
{\kern0.4\wd0\vrule height0.9\ht0\hss}\box0}}}}
\def\Zl{{\mathchoice
{\setbox0=\hbox{$\displaystyle\rm Z$}\hbox{\hbox to0pt
{\kern0.4\wd0\vrule height0.9\ht0\hss}\box0}}
{\setbox0=\hbox{$\textstyle\rm Z$}\hbox{\hbox to0pt
{\kern0.4\wd0\vrule height0.9\ht0\hss}\box0}}
{\setbox0=\hbox{$\scriptstyle\rm Z$}\hbox{\hbox to0pt
{\kern0.4\wd0\vrule height0.9\ht0\hss}\box0}}
{\setbox0=\hbox{$\scriptscriptstyle\rm Z$}\hbox{\hbox to0pt
{\kern0.4\wd0\vrule height0.9\ht0\hss}\box0}}}}
\def\Ql{{\mathchoice
{\setbox0=\hbox{$\displaystyle\rm Q$}\hbox{\hbox to0pt
{\kern0.4\wd0\vrule height0.9\ht0\hss}\box0}}
{\setbox0=\hbox{$\textstyle\rm Q$}\hbox{\hbox to0pt
{\kern0.4\wd0\vrule height0.9\ht0\hss}\box0}}
{\setbox0=\hbox{$\scriptstyle\rm Q$}\hbox{\hbox to0pt
{\kern0.4\wd0\vrule height0.9\ht0\hss}\box0}}
{\setbox0=\hbox{$\scriptscriptstyle\rm Q$}\hbox{\hbox to0pt
{\kern0.4\wd0\vrule height0.9\ht0\hss}\box0}}}}
\def\Rl{{\mathchoice
{\setbox0=\hbox{$\displaystyle\rm R$}\hbox{\hbox to0pt
{\kern0.4\wd0\vrule height0.9\ht0\hss}\box0}}
{\setbox0=\hbox{$\textstyle\rm R$}\hbox{\hbox to0pt
{\kern0.4\wd0\vrule height0.9\ht0\hss}\box0}}
{\setbox0=\hbox{$\scriptstyle\rm R$}\hbox{\hbox to0pt
{\kern0.4\wd0\vrule height0.9\ht0\hss}\box0}}
{\setbox0=\hbox{$\scriptscriptstyle\rm R$}\hbox{\hbox to0pt
{\kern0.4\wd0\vrule height0.9\ht0\hss}\box0}}}}
\def\Cl{{\mathchoice
{\setbox0=\hbox{$\displaystyle\rm C$}\hbox{\hbox to0pt
{\kern0.4\wd0\vrule height0.9\ht0\hss}\box0}}
{\setbox0=\hbox{$\textstyle\rm C$}\hbox{\hbox to0pt
{\kern0.4\wd0\vrule height0.9\ht0\hss}\box0}}
{\setbox0=\hbox{$\scriptstyle\rm C$}\hbox{\hbox to0pt
{\kern0.4\wd0\vrule height0.9\ht0\hss}\box0}}
{\setbox0=\hbox{$\scriptscriptstyle\rm C$}\hbox{\hbox to0pt
{\kern0.4\wd0\vrule height0.9\ht0\hss}\box0}}}}
\def\Hl{{\mathchoice
{\setbox0=\hbox{$\displaystyle\rm H$}\hbox{\hbox to0pt
{\kern0.4\wd0\vrule height0.9\ht0\hss}\box0}}
{\setbox0=\hbox{$\textstyle\rm H$}\hbox{\hbox to0pt
{\kern0.4\wd0\vrule height0.9\ht0\hss}\box0}}
{\setbox0=\hbox{$\scriptstyle\rm H$}\hbox{\hbox to0pt
{\kern0.4\wd0\vrule height0.9\ht0\hss}\box0}}
{\setbox0=\hbox{$\scriptscriptstyle\rm H$}\hbox{\hbox to0pt
{\kern0.4\wd0\vrule height0.9\ht0\hss}\box0}}}}
\def\Ol{{\mathchoice
{\setbox0=\hbox{$\displaystyle\rm O$}\hbox{\hbox to0pt
{\kern0.4\wd0\vrule height0.9\ht0\hss}\box0}}
{\setbox0=\hbox{$\textstyle\rm O$}\hbox{\hbox to0pt
{\kern0.4\wd0\vrule height0.9\ht0\hss}\box0}}
{\setbox0=\hbox{$\scriptstyle\rm O$}\hbox{\hbox to0pt
{\kern0.4\wd0\vrule height0.9\ht0\hss}\box0}}
{\setbox0=\hbox{$\scriptscriptstyle\rm O$}\hbox{\hbox to0pt
{\kern0.4\wd0\vrule height0.9\ht0\hss}\box0}}}}
\newcommand{\ch}{\mathcal H}

  \newcommand{\Fd}{\mathfrak{D}}

\newcommand{\fl}{\mathfrak{l}}  \newcommand{\Fl}{\mathfrak{L}}

\newcommand{\ft}{\mathfrak{t}}

\newcommand{\eps}{\epsilon}

\DeclareMathOperator{\MC}{\boldsymbol{\mathsf{M}}}
\DeclareMathOperator{\MCO}{\boldsymbol{\widehat{\mathsf{M}}}}

\DeclareMathOperator{\MCOW}{\rm Master\;\;Constraint\;\;
Operator}
\DeclareMathOperator{\MCPW}{\rm Master\;\;Constraint\;\;Programme}

\title{Testing the\\ $\MCPW$\\ for Loop Quantum Gravity\\ 
II. Finite Dimensional Systems}
\author{B.
Dittrich\thanks{dittrich@aei.mpg.de, bdittrich@perimeterinstitute.ca},
T. Thiemann\thanks{thiemann@aei.mpg.de, tthiemann@perimeterinstitute.ca}\\
\\
Albert Einstein Institut, MPI f. Gravitationsphysik\\
Am M\"uhlenberg 1, 14476 Potsdam, Germany\\
\\
and\\
\\
Perimeter Institute for Theoretical Physics \\
31 Caroline Street North, Waterloo, ON N2L 2Y5, Canada}
\date{{\small Preprint AEI-2004-117}}
\begin{document}

\maketitle

\begin{abstract}
This is the second paper in our series of five in which we test the Master 
Constraint Programme for solving the Hamiltonian constraint in Loop 
Quantum Gravity. In this work we begin with the simplest examples:
Finite dimensional models with a finite number of first or second class 
constraints, Abelean or non -- Abelean, with or without structure 
functions.
 \end{abstract}

\newpage

\tableofcontents

\section{Introduction}
\label{s1}

We continue our test of the Master Constraint Programme 
\cite{7.0} for Loop Quantum Gravity (LQG) \cite{1.1,7.2,7.3}
which we started in the companion paper \cite{I} and will continue
in \cite{III,IV,V}. The Master Constraint Programme is a new idea to 
improve on 
the current situation with the Hamiltonian constraint operator for LQG
\cite{7.1}. In short, progress on the solution of the Hamiltonian 
constraint has been slow because of a technical reason: the Hamiltonian 
constraints 
themselves are not spatially diffeomorphism invariant. This means that one 
cannot first solve the spatial diffeomorphism constraints and then 
the Hamiltonian constraints because the latter do not preserve the 
space of solutions to the spatial diffeomorphism constraint
\cite{8.2}. On the other hand, the space of solutions  
to the spatial diffeomorphism constraint \cite{8.2} is relatively easy to 
construct starting from the spatially diffeomorphism invariant 
representations on which LQG is based \cite{7.4} which are therefore 
very natural to use and, moreover, essentially unique. Therefore one would 
really like to keep these structures. The Master Constraint 
Programme removes that technical obstacle by replacing the Hamiltonian 
constraints by a single Master Constraint which is a spatially 
diffeomorphism invariant integral of squares of the individual 
Hamiltonian constraints which encodes all the necessary information about 
the constraint surface and the associated invariants. See e.g. 
\cite{7.0,I} for a full discussion of 
these issues. Notice that the idea of squaring constraints is not new,
see e.g. \cite{2.1}, however, our concrete implementation is new and also 
the Direct Integral Decomposition (DID) method for solving them, see 
\cite{7.0,I} for all the details. 

The Master Constraint for four 
dimensional General Relativity will appear in \cite{8.1} but before we 
test its semiclassical limit, e.g. using the methods of \cite{8.3,8.5} 
and try to solve it by DID methods we want to test the programme in the  
series of papers \cite{I,III,IV,V}. 
We begin by studying finite dimensional systems, in particular:
a finite number of Abelean constraints linear in the momenta which will 
also play an important role for \cite{IV}, a system with second class 
constraints, first class constraints with structure constants at most 
quadratic in the momenta and first class constraints linear momenta with 
structure functions.

\section{Finite Number of Abelean First Class Constraints Linear in the 
Momenta}
\label{s3}

 Our first example is a system with configuration manifold $\Rl^n$, 
coordinatized by $x^i,i=1,\ldots,n$ and $m<n$ commuting constraints
\be
C_i=p_i \q i=1,\ldots,m \q,
\ee
where the $p_i$'s are the conjugated momenta to the $x^i$'s.

All phase space functions which do not depend on $x_i$ are Dirac observables, i.e. functions which commute with the constraints (on the constraint hypersurface). A Dirac observable which depends on $p_i,\,i=1,\ldots,m$ is equivalent to the Dirac observable obtained from the first one by setting $p_i=0$ (since these two observables will coincide on the constraint hypersurface). Therefore it is sufficient to consider observables which are independent of the first $m$ configuration observables and of the first $m$ conjugated momenta. A canonical choise for an observable algebra is the one generated by $x^i,p_i,\,i=(m+1),\ldots,n$.

To quantize the system, we will start with an auxilary Hilbert space $L_2(\Rl^n)$ on which the operators $\hat x^i$ act as multiplication operators and the momenta as derivatives, i.e. we use the standard Schr\"odinger representation:
\ba 
\hat x^i \,\psi(\bold x)=x^i  \,\psi(\bold x) \q\q\q  \hat p_i\psi(\bold x)=-i\hbar \,\partial_i  \psi(\bold x) \q.
\ea

According to the Master Constraint Programme, we have to consider the spectral resolution of the Master Constraint
\be
\MCO=-\hbar^2\,\sum_{i=1}^m \partial_i^2=:-\hbar^2 \Delta_m \q.
\ee
This spectral resolution can be constructed with the help of the spectral resolutions of the operators $\hat p_i=-i\hbar\partial_i$, which are well known to be given by the Fourier transform:
\be
\psi(\bold x)= \int_{\Rl}  <b_{k_i},\psi(x^1,\ldots,x^{i-1},\cdot,x^{i+1},x^n) >_i b_{k_i}(x^i) dk_i
\ee
with generalized eigenfunctions
\ba
&& b_{k_i}(x^i)=\tfrac{1}{\sqrt{2\pi}}\exp(ik_i x^i) \q\q\text{and} \nn \\
&& <b_{k_i},\psi(x^1,\ldots,x^{i-1},\cdot,x^{i+1},x^n )>_i=\int_{\Rl} \tfrac{1}{\sqrt{2\pi}}\exp(-ik_i x^i)\psi(\bold x) dx^i 
\ea
and $\psi \in L_2(\Rl^n)$. The spectrum of $\hat p_i$ is therefore $\text{spec}(\hat p_i)=\Rl$. Moreover the spectral projectors of $\hat p_i$ and $\hat p_j$ commute, so we can achieve a simultaneous diagonalization of all the $C_i$'s (that is the Fourier transform with respect to $(x^1,\ldots,x^m)$:
\ba \label{specres3}
\psi(\bold x)=\int_{\Rl^m} <\prod_{i=1}^{m} b_{k_i}(\cdot),\psi(\cdots,x_{m+1},\ldots,x_n) >_{(m)} \,
 \bigg(\prod_{i=1}^{m} b_{k_i}(x^i) \bigg)     \,dk_1\cdots dk_m
\ea
with
\ba 
<\prod_{i=1}^{m} b_{k_i}(\cdot),\psi(\cdots,x_{m+1},\ldots,x_n)  >_{(m)}
=\int_{\Rl^m}\bigg(\prod_{i=1}^{m} b_{k_i}(x^i)\bigg)\psi(\bold x) dx^1 \cdots dx^m \q.
\ea
This decomposition of a function $\psi \in L_2(\Rl^n)$ corresponds to a decomposition of the Hilbert space $L_2(\Rl^n)$ into a direct integral of Hilbert spaces
\be
L_2(\Rl^n)\simeq\int_{\Rl^m}\ch_{(k_1,\ldots,k_m)} dk_1\cdots dk_m \q.
\ee
where each $\ch_{(k_1,\ldots,k_m)}$ is isomorphic to $L_2(\Rl^{n-m})$.

 Since the Master Constraint Operator is a polynomial of the $C_i$'s, 
 according to spectral calculus  we already achieved the spectral 
 resolution of the Master Constraint. The spectrum of $\MCO$ is given by 
$\text{spec}(\hat M)=\text{clos}\{ \sum_{i=1}^m k_i^2,\,k_i \in \Rl\}=\Rl_+$. The generalized eigenfunctions of $\hat M$ to the (generalized) eigenvalue $0$ can be read off from (\ref{specres3}) to be
\ba
\Psi(\bold x)&=&\frac{1}{(\sqrt{2\pi})^m}\exp(i\,\sum_{i=1}^m k_i x^i)_{|k_i=0} \psi(x^{m+1},\cdots,x^n) \nn \\
&=&\frac{1}{(\sqrt{2\pi})^m}\psi(x^{m+1},\cdots,x^n) \q \q \text{with}\q \psi \in L_2(\Rl^{n-m})
\ea
i.e. functions which do not depend on $(x^1,\ldots,x^m)$. So one could conclude that the physical Hilbert space is $\ch_{(0,\ldots,0)}\simeq L_2(\Rl^{n-m})$, a space of functions of $(x^{m+1},\ldots,x^n)$. The action of the elementary Dirac observables $\hat x^i$ and $\hat p_i$ for $i=(m+1),\ldots,n$ is well defined on this physical Hilbert space.

However we would like to go through the explicit procedure for constructing the direct integral decomposition for a seperable Hilbert space, since we will use it later on in another example. The general procedure is explained in detail in \cite{I}.

We will work in the fourier-transformed picture, i.e. the Hilbert space is $L_2(\Rl^n)$ as a function space of the momenta $(p_1,\ldots,p_n)$ and the Master Constraint Operator becomes the multiplication operator $\MCO=\sum_{i=1}^m p_i^2$.

To begin with we have to choose a set of orthonormal vectors $\{\Omega_i\}_i$ from which $\ch=L_2(\Rl^n)$ can be generated through repeated applications of the Master Constraint Operator $\MCO$. To this end we use the fact (see \cite{howe}), that
\be  \label{first:basis1}
\{\exp\!\big(-\tfrac{1}{2}\sum_{j=1}^m p_j^2 \big) \big(\sum_{j=1}^m p_j^2 \big)^k \psi_{s,l}\,|s,k \in \Nl, l \in \fl_s \}
\ee
is a basis for $L_2(\Rl^m)$. Here $\psi_{s,l}$ are harmonic 
polynomials\footnote{i.e. homogeneous polynomials annihilated by the Laplace operator on $\Rl^m$} of degree $s$ on $\Rl^m$, and the index $l$ takes values in some finite index set $\fl_s$, which depends on $m$ and $s$. Obviously the set (\ref{first:basis1}) can be generated by applying repeatedly $\MCO$ (more correctly $\tilde M$ where $\MCO=\tilde M \otimes 1_{L_2(\Rl^{m-n})}$, so that $\tilde M$ is an operator on $L_2(\Rl^m)$) from the set
\be  
\{\tilde \Omega_{sl}:=\exp\!\big(-\tfrac{1}{2}\sum_{j=1}^m p_j^2 \big) \psi_{s,l}\,|s \in \Nl, l \in \fl_s \} \q.
\ee

Moreover the spaces generated from different $\tilde \Omega_{s,l}$ are mutually orthogonal. (This can be seen if one introduces spherical coordinates. Then the restrictions of the harmonic polynomials to the unit sphere, which are given by the spherical harmonics, are orthogonal in $L_2(S^{m-1},d\omega_{n-1})$ where $d\omega_{n-1}$ is the uniform measure on the sphere $S^{m-1}$.) 
So we conclude that the set of orthonormal vectors
\be
\{ \Omega_{slt}:=N_{sl}\exp\!\big(-\tfrac{1}{2}\sum_{j=1}^m p_j^2 \big)  \psi_{s,l} \otimes \phi_t (p_{m+1},\ldots ,p_n) \,|s \in \Nl, l \in \fl_s, t \in \ft \} \q.
\ee
where $N_{sl}$ are normalization constants and $\{\phi_t\}_{t\in \ft}$ is an orthonormal basis of $L_2(\Rl^{n-m})$ fulfills all the demands required above, i.e. 
\be 
L_2(\Rl^n)=\sum_{s,l,t}{}^\oplus \overline{\text{span}\{p(\MCO)\Omega_{slt}|p\q \mbox{polynomial}\}} \q.
\ee
 Now the vectors $\Omega_{slt}$ are tensor products of the form 
 $\Omega_{slt}=\Omega_{sl}\otimes \phi_t$ where $\Omega_{sl} \in 
\ch_1:=L_2(\Rl^m)$ and $\phi_t \in \ch_2:=L_2(\Rl^{n-m})$ and furthermore 
$\MCO=\tilde M \otimes 1_{\ch_2}$ acts only on the first factor. Hence we have for the spectral measures $\mu_{slt}(\lambda)$:
\ba \label{specmeassym1}
\mu_{slt}&:=&
<\Omega_{sl}\otimes\phi_t,(\Theta(\lambda-\tilde M)\Omega_{sl})
\otimes\phi_t>_\ch= 
<\Omega_{sl},\Theta(\lambda-\tilde{M})\Omega_{sl}>_{\ch_1}||\phi_t||_{\ch_2} 
\nn \\
&=&<\Omega_{sl},\Theta(\lambda-\tilde{M})\Omega_{sl}>_{\ch_1}=:\mu_{sl} \q 
,
\ea
so that we only need to consider the spectral measures $\mu_{sl}$ with 
respect to  the operator $\tilde M$ on $\ch_1$. Additionally, using the 
rotational symmetry of the Master Constraint Operator (and an idea outlined 
in \cite{howe}) one can simplify the calculations even more: The space of 
harmonic polynomials on $\Rl^m$ of degree $s$ is an irreducible module for 
the rotation group $O(m)$ under the left regular representation, which acts as
\be
U(R): \psi(\vec p)\mapsto \psi(R^{-1}\cdot \vec p)
\ee
on the space of functions over $\Rl^m$. Here $R$ is a rotation matrix for $\Rl^m$. This representation is unitary if considered on the Hilbert space $\ch_1=L_2(\Rl^m)$. The irreducibilty of the representation ensures that one can generate all vectors $\{\Omega_{sl}\}_{l\in \fl}$ by applying rotations $U(R)$ to just one vector, say $\Omega_{s}:=\Omega_{s0}$. But these rotations leave the spectral projectors of $\tilde M$ invariant (since $\tilde M$ commutes with the rotations $U(R)$), so that we have
\ba
\mu_{sl}&=&<\Omega_{sl},\Theta(\lambda-\tilde M) \Omega_{sl}> =<U(R)\Omega_{s0},\Theta(\lambda-\tilde M) U(R)\Omega_{s0}> \nn \\ 
&=&<\Omega_{s0},\Theta(\lambda-\tilde M) \Omega_{s0}>=\mu_{s0}=:\mu_s
\ea 
for an appropriate rotation $R$. Here one can see, that it is very helpful to know the symmetries of the Master Constraint Operator, i.e. unitary operators commuting with $\MCO$.  These may come from (exponentiated) strong Dirac observables, that is operators which commute with all the constraints on the whole Hilbert space $\ch$. Examples for these are operators of the type $1_{\ch_1} \otimes U_2$, i.e. unitary operators which act only on $\ch_2=L_2(\Rl^{n-m})$. We used this kind of symmetry in (\ref{specmeassym1}). However the $U(R)$'s from above are not of this type: The classical counterparts of their generators (i.e. angular momentum on $\Rl^m$) Poisson-commutes with the Master Constraint but they also vanish on the constraint hypersurface $\{p_i=0, i\leq m\}$. Nevertheless the rotation group $O(m)$ is useful in constructing the direct Hilbert space decomposition. Later we will see, that the vanishing of its (classical) generators on the constraint hypersurface corresponds to the fact, that the representation of the rotation group on the induced Hilbert space is trivial.

So we just need to select for each $s \in \Nl$ a particular homogeneous harmonic polynomial $\psi_s$ of degree s. We choose
\be
\psi_{s}(p_1,\ldots,p_m)=(p_1+ip_2)^s\q;\,\, s=0,1,2,\ldots
\ee
so that the vectors $\Omega_s$ become
\be
\Omega_s=\frac{1}{\sqrt{\pi^{m/2} s!}}e^{-\tfrac{1}{2}r_m^2}r_2^s e^{is\varphi}
\ee
where we introduced the coordinates $r_m=\sum_{i=1}^m p_i^2$,\,\,$r_2^2=p_1^2+p_2^2$\, and the angle $\varphi=\arctan(p_1/p_2)$. With this at hand we can compute the spectral measures $\mu_s$ (we will assume that $m\geq 4$):
\ba \label{firstlongcalc1}
\mu_s(\lambda)&=&<\Omega_s,\Theta(\lambda-\tilde M)\Omega_s>_{\ch_1} \nn \\
 &=& \frac{1}{\pi^{m/2}\,s!}\int_{\Rl^m} d^m p \,\Theta(\lambda-r^2_m) e^{-r^2_m} r_2^{2s} \nn\\
&\underset{r_m^2=r_2^2+r^2_{m-2}}{=}&\frac{1}{\pi^{m/2}\,s!}\int_0^{2\pi} d\varphi\int_{\Rl_+} dr_2 \int_{S^{m-3}} d\omega_{m-3} \int_{\Rl_+} dr_{m-2} \Theta(\lambda-r_m^2)r_2^{2s+1} r_{m-2}^{m-3} e^{-r^2_m} \nn \\
&\underset{\stackrel{r_2=r_m\cos \phi}{r_{m-2}=r_m \sin \phi}}{=}& \frac{2\pi \text{Vol}(S^{m-3})}{\pi^{m/2}\,s!}\int_0^{\pi/2} d\phi \int_{\Rl_+} dr_m\, \Theta(\lambda-r_m^2)r_m^{2s+m-1} (\cos \phi)^{2s+1} (\sin \phi)^{m-3} e^{-r_m^2} \nn \\
&\underset{y=r_m^2}{=}& \frac{2\pi \text{Vol}(S^{m-3})}{\pi^{m/2}\,s!} \,\frac{s!\,\, \Gamma(m/2 -1)}{2\,\Gamma(s+m/2)} \frac{1}{2}\int_{\Rl_+} dy\, \Theta(\lambda-y) \, y^{s+m/2-1} e^{-y}  \nn \\
&=&\frac{1}{\Gamma(s+m/2)}\int_{\Rl_+} dy\, \Theta(\lambda-y) \, y^{s+m/2-1} e^{-y}
\ea
%\mu_s(\infty)=1
From the second to the third line we transformed the cartesian coordinates 
$(p_3,\ldots, p_m)$ to spherical coordinates on $\Rl^{m-2}$ with radial coordinate $r_{m-2}^2=\sum_{i=3}^m p_i^2$ and angels varying over $S^{m-3}$. $\text{Vol}(S^{m-3})$ is the euclidian volume of the $(m-3)$-dimensional unit sphere. From the third to the fourth line we again introduced spherical coordinates $(r_m,\phi)$ for the two radii $r_2$ and $r_{m-2}$, so that $r_m^2=r_2^2+r^2_{m-2}$. Since $r_2,r_{m-2}$ are positive, $\phi$ varies over $[0,\pi/2]$. Then we integrated over $\phi$ and used a new integration variable $y=r^2_m$ for the remaining integral. In the last step we used that $\text{Vol}(S^{m-3})=2\pi^{m/2-1}/\Gamma(m/2-1)$.

To obtain the final spectral measure $\mu(\lambda)$ we have to sum over all the measures $\mu_{slt}\equiv \mu_{s}$ multiplied by constants $\alpha_{slt}$. We choose the constants such that $\sum_{l,t} \alpha_{slt}=2^{-s-1}$, so that the sum over all the $\alpha_{slt}$'s is one. The spectral measure becomes
\ba \label{firstlongcalc2}
\mu(\lambda)&=&\sum_{s,l,t} \alpha_{slt}\,\mu_{slt}(\lambda)=\sum_{s=0}^\infty 2^{-s-1} \mu_s(\lambda) \nn \\
    &=&\frac{2\pi \text{Vol}(S^{m-3})}{2\,\pi^{m/2}} \sum_{s=0}^\infty \frac{2^{-s}}{s!}\int_{\Rl_+} dr_2\int_{\Rl_+} dr_{m-2} \Theta(\lambda-r_m^2)r_2^{2s+1} r_{m-2}^{m-3} e^{-r^2_m} \nn \\
     &=&\frac{\pi \text{Vol}(S^{m-3})}{\pi^{m/2}}\int_{\Rl_+} dr_2\int_{\Rl_+} dr_{m-2} \Theta(\lambda-r_m^2)r_2\,e^{\tfrac{1}{2}r_2^2}r_{m-2}^{m-3} e^{-r^2_m} \nn \\
    &\underset{x=r_2^2/2}{=}&\frac{\pi \text{Vol}(S^{m-3})}{\pi^{m/2}}\int_{\Rl_+} dx\int_{\Rl_+} dr_{m-2} \Theta(\lambda-r_{m-2}^2-2x)e^{-x}e^{-r_{m-2}^2}r^{m-3}_{m-2} \nn \\
    &=&\frac{\pi \text{Vol}(S^{m-3})}{ \pi^{m/2}}\int_{\Rl_+} dr_{m-2}\Theta(\lambda-r_{m-2}^2)\big(1-e^{-\tfrac{1}{2}{\lambda}+\tfrac{1}{2}r^2_{m-2}}\big)e^{-r_{m-2}^2} r_{m-2}^{m-3} \nn \\
   &\underset{y=r^2_{m-2}}{=}&\frac{1}{\Gamma(m/2-1)}\int_{\Rl_+}dy\, \Theta(\lambda-y) \big(1-e^{-\tfrac{1}{2}\lambda+\tfrac{1}{2}y}\big)e^{-y}y^{m/2-2}
\ea 
%\mu_(\infty)=1
Here we inserted for $\mu_s$ the third line of (\ref{firstlongcalc1}), 
exchanged integration and summation, performed a variable transformation 
$x=r_2^2/2$ in the fourth line, and integrated over the new variable $x$ 
in the fifth line. Finally we changed to the integration variable 
$y=r^2_{m-2}$ in the last line and used the explicit expression 
$2\pi^{m/2-1}/\Gamma(m/2-1)$ for $\text{Vol}(S^{m-3}) $. This gives for 
the derivative of $\mu(\lambda)$:
\ba
\frac{d\mu(\lambda)}{d\lambda}&=&\frac{1}{\Gamma(m/2-1)}\int_{\Rl_+}dy\, \delta(\lambda-y) \big(1-e^{-\tfrac{1}{2}\lambda+\tfrac{1}{2}y}\big)e^{-y}y^{m/2-2} \nn \\
&&+\frac{1}{\Gamma(m/2-1)}\int_{\Rl_+}dy\, \Theta(\lambda-y) \big(\tfrac{1}{2}e^{-\tfrac{1}{2}\lambda+\tfrac{1}{2}y}\big)e^{-y}y^{m/2-2}\nn\\
&=&\frac{1}{2\,\Gamma(m/2-1)}\, e^{-\tfrac{1}{2}\lambda}\int_{\Rl_+}dy\, \Theta(\lambda-y)e^{-\tfrac{1}{2}y}y^{m/2-2}\q.
\ea
Next we have to calculate the Radon-Nikodym derivatives 
$\rho_s=\rho_{slt}$ of $\mu_s$ with respect to $\mu$. Since both measures 
are absolutely continuous with respect to the Lebesgue measure, we can write
\ba
\rho_s(\lambda)=\frac{d\mu_s(\lambda)}{d\mu(\lambda)}=
\frac{d\mu_s(\lambda)/d\lambda}{d\mu(\lambda)/d\lambda}=\frac{ 2 \,\Gamma(m/2 -1)}{\Gamma(m/2+s)}  \lambda^{s+m/2-1} e^{-\frac{1}{2}\lambda} \bigg[\int_0^\lambda dy\,  e^{-\tfrac{1}{2}y}y^{m/2-2}\bigg]^{-1}
\ea
For $0 <\lambda \leq \infty $  the derivatives $\rho_s(\lambda)$ will be 
some strictly positive numbers for all $s \in \Nl$. But for the limit $\lambda \rightarrow 0$ we apply L'hospital's rule, resulting in
\ba
\underset{\lambda \rightarrow 0}{\text{lim}}\,\rho_s(\lambda)=\frac{ 2\,\Gamma(m/2 -1)}{\Gamma(m/2+s)} \underset{\lambda \rightarrow 0}{\text{lim}}\, \frac{(s+m/2-1)\lambda^{s+m/2-2}
-\tfrac{1}{2}\lambda^{s+m/2-1}
}{\lambda^{m/2-2}}
\ea
which gives $\rho_s(0)=0$ for $s>0$ and $\rho_0(0)=2$.

Now we can construct the induced Hilbert space $\ch^\oplus(0)$. We notize that there is only one linearly independent harmonic polynomial of degree zero, namely $\psi_{00}\equiv 1$. Therefore $\ch^\oplus(0)$ has an orthonormal basis $\{e_t\}_{t\in \ft}$ which corresponds to the set of vectors $\{\Omega_{00t}\}_{t\in \ft}$. Hence we can identify $\ch^\oplus(0)$ with the Hilbert space $L_2(\Rl^{n-m})$ of functions in the variables $(p_{m+1},\ldots,p_n)$. Interestingly, because of $\rho_{slt}(\lambda)>0$ for $\lambda>0$, the induced Hilbert spaces $\ch^\oplus(\lambda)$ for $\lambda >0$ are in some sense much bigger, since they have a basis $\{e_{slt}|s\in \Nl,\, l \in \fl,\,t \in \ft \}$ corresponding to all the vectors $\Omega_{slt}$. The latter can be seen as a basis in $L_2(S^{m-1}, \lambda^\frac{m-1}{2}d\omega_{m-1} )\otimes L_2(\Rl^{n-m})$, therefore we can identify $\ch^\oplus(\lambda)$ with $L_2(S^{m-1}, \lambda^\frac{m-1}{2}d\omega_{m-1} )\otimes L_2(\Rl^{n-m})$ for $\lambda >0$.

 We would like to mention, that the space of (formal) solutions to the 
 Master Constraint Operator is much bigger than the space of functions, 
 which are independent of the first $m$ coordinates. For instance, 
 functions, harmonic in the first $m$ coordinates, are solutions of the 
 master constraint but they are unphysical because they do not solve the 
 individual constraints $C_i$. These solutions automatically do not appear 
in the spectral resolution of the Master Constraint. The intuitive reason 
for this will be discussed in the conclusions.
%This shows that the 
%spectral analysis of the master constraint selects the correct solution 
%space (at least for this example).   

\section{A Second Class System }
\label{s4}

Here we will discuss a simple second class system, given by the constraints
\ba
C_i=x^i \q\q\text{and}\q\q  B_i=p_i  \q\q i=1,\ldots,m \nn \\
\ea
on the phase space $\Rl^{2n}$. (The $x^i$ denote configuration variables and the $p_i$ their conjugated momenta.) The Poisson brackets are given by $\{C_i,B_j\}=\delta_{ij}$ and all other Poisson brackets vanish. Because of the Heisenberg Uncertainty relations one cannot expect to find eigenfunctions of the Master Constraint Operator corresponding to the eigenvalue zero. We will therefore alter the Master Constraint and validate whether this gives sensible results. 

A complete set of observables is given by phase space functions which are independent on the first $m$ configuration variables and momenta.    

As in the last case we will quantize the system by choosing the auxilary Hilbert space $\Fl^2(\Rl^n)$ of functions of the configuration variables $\psi(\bold x)$. The operators $\hat x^i$ act as multiplication operators and the momentum operators $p_i$ as derivatives. 

The Master Constraint Operator is defined as the sum of the squares of the constraints but we can as well consider the following slight variation of this prescription:
\ba \label{0.100}
\MCO =\sum_{i=1}^m \tfrac{1}{2}(%m_i^{-1}
 \hat p_i^2 + %m_i
\omega_i^2 (\hat x^i)^2)\q,
\ea
where 
%$m_i$ and
 $\omega_i$ are $
%2
m$ positive constants. This Master Constraint coincides with the 
Hamiltonian for $m$ uncoupled harmonic oscillators with different frequencies. This Hamiltonian has pure point and positive spectrum, the lowest eigenvalue being $\lambda_0=\tfrac{\hbar}{2}\sum_{i=1}^m \omega_i$.  

 Since zero is not in the spectrum of the Master Constraint Operator we 
 have to alter $\MCO$, such that its spectrum includes zero. Of course we 
 have to check in the end, whether this procedure gives a sensible quantum 
 theory. The simplest thing one can do, is to subtract $\lambda_0$ from 
 the Master Constraint to obtain $\MCO'=\MCO-\lambda_0$. This is 
 equivalent to the normal ordering of the Master Constraint Operator.
This could be done also for the limit $m\to \infty$ if we multiply 
the individual operators in the sum of (\ref{0.100}) by positive constants 
$Q_m$ with $\sum_m Q_m<\infty$. This condition will naturally reappear
in the free field theory examples of \cite{IV}.  
% (i.e. one introduces annihilation and creation operators and writes all 
%annihilation operators (by hand) to the right of the creation operators). 

Now the spectrum of $\MCO'$ includes zero and we can construct the induced Hilbert space to the eigenvalue zero. To this end we have to find a cyclic basis, which we will choose to be a tensor basis of the following kind:
\be
\Omega_{n_1,\ldots,n_m,k}= f_{n_1,\ldots,n_m}(x_1,\ldots, x_m)\otimes h_k(x_{m+1}, \ldots, x_n)
\ee
where $f_{n_1,\ldots,n_m},\, n_i\in \Nl$ are the (normalized) eigenfunctions of $\MCO$ (considered on $L_2(\Rl^m)$) and $\{h_k\,|\,k\in \Nl\}$ is an orthonormal basis in $L_2(\Rl^{n-m})$. The application of $\MCO$ to the vectors $\Omega_{n_1,\ldots,n_m,k}$ just multiplies them with the corresponding eigenvalue so that we need them all in our cyclic basis. The associated spectral measures are
\be
\mu_{n_1,\ldots,n_m,k}(\lambda)=<\Omega_{n_1,\ldots,n_m,k},\,\Theta(\lambda-\MCO')\Omega_{n_1,\ldots,n_m,k}>=\Theta(\lambda-\lambda_{n_1,\ldots,n_m})
\ee
 where $\lambda_{n_1,\ldots,n_m}=\tfrac{\hbar}{2}\sum_{i=1}^m n_i 
\,\omega_i$. Only measures with $n_1=n_2=\ldots=n_m=0$ have the point zero in their support. The final spectral measure can be defined to be
\be
\mu(\lambda)=\sum_{n_1,\ldots,n_m,k \in \Nl} 
2^{-(n_1+1)}\cdots 2^{-(n_m+1)}\cdot 
2^{-(k+1)}\mu_{n_1,\ldots,n_m,k}(\lambda)=\Theta(\lambda)+ \nu(\lambda)     
\ee
where $\nu(\lambda)=\sum_{n_1,\ldots,n_m\geq 1,k \in \Nl} 
2^{-(n_1+1)}\cdots 2^{-(n_m+1)}\cdot 
2^{-(k+1)}\mu_{n_1,\ldots,n_m,k}(\lambda) $ 
is a pure point measure which does not have support at zero. The relevant Radon-Nikodym derivatives for the induced Hilbert space to the eigenvalue zero are therefore
\be
\rho_{n_1=0,\ldots,n_m=0,k}(0)=1  \q,
\ee 
hence we can interpret the set $\{\Omega_{n_1=0,\ldots,n_m=0,k}\,|\,k\in \Nl\}$ as a basis in the induced Hilbert space. So as expected for a pure point spectrum the induced Hilbert space is just the (proper) eigenspace to the eigenvalue zero. This eigenspace can be identified with the space $L_2(\Rl^{n-m})$ by mapping $\Omega_{n_1=0,\ldots,n_m=0,k}$ to the basis vector $h_k(x_{m+1}, \ldots, x_n)$ in $L_2(\Rl^{n-m})$. Also the action of Dirac observables (that is quantized phase space functions, which do not depend on the first $m$ configuration variables and momenta) coincides on the null eigenspace of $\MCO'$ and on $L_2(\Rl^{n-m})$. We can therefore conclude, that the physical Hilbert space is given by $L_2(\Rl^{n-m})$, which is the result one would expect beforehand.

\section{Finite Number of First Class Constraints at most Quadratic in the 
Momenta}
\label{s5}

 If the constraints generate a (semi-simple) compact Lie group it is in 
 general straightforward to apply the Master Constraint Programme. The 
 Master Constraint Operator coincides in this case with the Casimir of the 
 Lie group and has a pure point spectrum. The direct integral 
 decomposition of the kinematical Hilbert space (which in this case is 
 truly a direct sum decomposition) is equivalent to the reduction of the 
 given representation of the Lie group into irreducible representations 
 and the physical Hilbert space corresponds to the isotypical 
component\footnote{Compact Lie groups are completely reducible. The 
isotypical component of a given equivalence class within a reducible 
representation is the direct sum of all its irreducible representations
into which it can be decomposed and which lie in the given equialence 
class.} of 
 the equivalence class of the trivial representation.

\subsection{$SU(2)$ Model with Compact Gauge Orbits}
\label{s5.1}

Here we consider the configuration space $\Rl^3$ with the three $so(3)$-generators as constraints:
\ba
L_i=\eps_{\,\, ij}^k x^j p_k, \q   \quad \{L_i,L_j\}=\eps_{\,\, ij}^k L_k
% L_1=x^2p_3-x^3 p_2 \\
% L_2= x^3p_1-x^1p_3 \\
%L_3=x^1p_2-x^2 p_1
\ea
where $\eps_{\,\, ij}^k=\eps_{ijk}$ is totally antisymmetric with $\eps_{123}=1$ and we summed over repeated indices. (In the following, indices will be raised and lowered with respect to the metric $g_{ik}=\text{diag}(+1,+1,+1)$.)

The observable algebra of this system is generated by
\ba \label{obs}
d=x^ip_i \quad \q\q\q e^+=x^ix_i \quad \q\q\q e^-=p^ip_i  \nn \\ 
\{d,e^\pm\}=\mp 2 e^\pm \q\q\q\q \q\q \q \quad \{e^+,e^-\}=4d \q.
\ea 
It constitutes an $sl(2,\Rl)$-algebra. We have the identity
\be \label{classid}
-d^2+e^+ e^-=L_iL^i
\ee
between the Casimirs of the constraint and observable algebra. 

\subsubsection{Quantization}

We start with the auxilary Hilbert space $\Fl^2(\Rl^3)$ of square integrable functions of the coordinates. The momentum operators are $\hat p_j=-i(\hbar)\partial_j$ and the $\hat x^j$ act as multiplication operators. There arises no factor ordering ambiguity for the quantization of the constraints, but for the observable algebra to close, we have to choose:
\ba \label{sl2rrep}
\hat d= \tfrac{1}{2}(\hat x^i\hat p_i+\hat p_i \hat x^i)=\hat x^i  \hat p_i - \tfrac{3}{2}i \hbar \q\q\q\q
\hat e^+=\hat x^i \hat x_i \quad \q\q\q \hat e^-=\hat p^i \hat p_i 
\ea
The commutators between constraints and between observables are then obtained by replacing the Poisson bracket with $\tfrac{1}{i\hbar}\big[\cdot,\cdot\big]$.

The identity (\ref{classid}) is altered to:
\ba \label{quid}
\hat d^2-\tfrac{1}{2}(\hat e^+ \hat e^-+ \hat e^- \hat e^+)-\tfrac{3}{4}\hbar^2 = -\hat L_i \hat L^i \q.
\ea

For the implementation of the Master Constraint Progamme we have to construct the direct integral decomposition with respect to the Master Constraint Operator
\be
\MCO:=\hat L_1^2+ \hat L_2^2+ \hat L_3^2=-\bigg(\hat d^2-\tfrac{1}{2}( \hat e^+  \hat e^-+  \hat e^-  \hat e^+)-\tfrac{3}{4} \hbar^2\bigg)
\ee
 The Master Constraint Operator is the Casimir of $SO(3)$ on $\Fl^2(\Rl^3)$. Its spectrum and its (normalized) eigenfunction are well known, the latter are given by the spherical harmonics. To discuss these we make a coordinate transformation to spherical coordinates:
\ba
x_1&=&r \cos(\phi)\sin(\theta)    \q \phi \in \big[0,2\pi)\; \text{and}\; \theta\in \big[0,\pi) \nn \\
x_2&=&r \sin(\phi)\sin(\theta) \nn \\
x_3&=&r \cos(\theta)  \nn \\ 
\\
d^3x&=&r^2 \sin(\theta)dr\,d\theta \,d\phi \nn \\
\widehat{L^2}&=&\hat L_1^2+ \hat L_2^2+ \hat L_3^2=-\hbar^2\Big[ \frac{1}{\sin^2\theta} \partial_\phi^2+\frac{1}{\sin \theta}\partial_\theta (\sin \theta \,\partial_\theta)\Big]  \, \\
\hat p^2&=&-\hbar^2 \Delta=-\hbar^2\Big(\frac{1}{r^2}\partial_r(r^2\,\partial_r)+\frac{1}{r^2\sin\theta}\partial_\theta(\sin\theta\partial_\theta)+\frac{1}{r^2\sin^2\theta}\partial^2_\phi\Big) \nn \\
&=& -\hbar^2 \frac{1}{r^2}\partial_r(r^2\,\partial_r) + \frac{1}{r^2}\widehat{L^2} \nn \\
\hat x^2&=&r^2   
\ea

The eigenfunctions of $\MCO=\widehat{L^2} $ to the eigenvalue $\hbar^2\,l(l+1), l \in \Nl$ are of the form $Y_{lm}(\theta,\phi)\,R(r)$ where $Y_{lm},\,l\in \Nl,-l\leq m \leq l $ are the spherical harmonics on the two-dimensional sphere $S^2$ and $R(r)$ is an arbitrary function in $L_2(\Rl_+, r^2dr)$. To discuss the direct integral decomposition of the kinematical Hilbert space $L_2(\Rl^3)$ we have to find a cyclic basis of this Hilbert space. We will choose the eigenbasis of $\hat{H}:=\hat{e}^-+\hat{e}^+$, i.e. the three-dimensional harmonic oscillator Hamiltonian. Its (normalized) eigenfunctions are given by {\cite{Galindo}
\be \label{SO(3)cyclic}
\psi_{nlm}(r,\theta,\phi)=N_{nl}\, r^l\, \exp(-\frac{r^2}{2 \hbar})\, M(-n, l+\tfrac{3}{2}, \tfrac{1}{\hbar}r^2)\, Y_{lm}(\theta,\phi)
\ee
 where the index $n$ takes values in $\Nl$, the constant $N_{nl}$ is a 
 normalization constant and $M(-n, l+\tfrac{3}{2}, r^2)$ are confluent 
 hypergeometric functions. Since the first argument of these is a whole 
 negative number, the functions $M(-n, l+\tfrac{3}{2}, r^2)$ reduce to 
 polynomials in $r^2$ of order $n$. Therefore, for fixed indices $m,l$ the 
 functions $\psi_{nlm}$ are polynomials in $r$ with minimal degree $r^l$. 
 (From this, one can see, that $L_2(\Rl^3)$ is not equivalent to a tensor 
 product $L_2(\Rl_+,r^2dr)\otimes L_2(S^2, \sin\theta d\theta d\phi)$. If 
 this would be the case the $\{\psi_{nlm}\,|\, n \in \Nl\}$ should span 
 the whole Hilbert space $L_2(\Rl_+,r^2dr)$ for arbitrary fixed indices 
 $m,l$. However, they span only the subspace of polynomials with minimal 
 degree $l$. We will come back to this point later on.) 

Applying $\MCO$ to the functions $\psi_{nlm}$ just multiplies them with $\hbar^2 l(l+1)$, hence we need them all as cyclic vectors $\Omega_{nlm}$. The associated spectral measures can easily be calculated to be
\be
\mu_{nlm}(\lambda)=<\psi_{nlm}, \Theta(\lambda-\MCO) \psi_{nlm}>=<\psi_{nlm}, \Theta(\lambda-\hbar^2l(l+1)) \psi_{nlm}>= \Theta(\lambda-\hbar^2l(l+1) ) \q .
\ee
The spectral measures are pure point, they are homogeneous in $n$ and $m$ and have only the point $\hbar^2 l(l+1)$ as support. Hence the kinematical Hilbert space decomposes into eigenspaces of $\MCO$ in the following way
\be
L_2(\Rl^3)=\sum_{l=0}^{\infty} \text{clos}(\text{span}\{\psi_{nlm}\,|\, n\in \Nl, -l\leq m \leq l \}) \q .
\ee
The induced Hilbert space to the eigenvalue $\lambda=0$ is given by  $\text{clos}(\text{span}\{\psi_{n00}\,|\, n\in \Nl\})$. These functions are constant in $\theta,\phi$ and are polynomials in $r^2$ of order $n$ (weighted with a gaussian factor). The above set can be taken as an orthonormal basis in $L_2(\Rl_+,r^2 dr)$, therefore we can identify the physical Hilbert space with $L_2(\Rl_+,r^2 dr)$.

On this (physical) Hilbert space the observable algebra is given by
\ba
\hat e^+=r^2 \q\q \hat e^- =-\hbar^2\frac{1}{r^2}\partial_r(r^2\,\partial_r) \q \q \hat d=-i\hbar\,(r\partial_r+\tfrac{3}{2})
\ea
and because of (\ref{quid}) we have
\be
\hat d^2-\tfrac{1}{2}(\hat e^+ \hat e^-+ \hat e^- \hat e^+)=\tfrac{3}{4}\hbar^2 \q .
\ee
%(This gives for the $sl(2,\Rl)$-Casimir $\Fc=\tfrac{1}{4} (\hat d^2-\tfrac{1}{2}(\hat e^+ \hat e^-+ \hat e^- \hat e^+))=\tfrac{3}{16}$.)

%\subsubsection{Spherical Harmonics}

%The spherical harmonics are defined for $l\in \Nl,\,-l\leq m\leq l$. For $m\geq0$ they are given by
%\ba
%Y_{lm}(\theta,\phi)=\sqrt{\frac{(2l+1)(l-m)!}{4\pi(l+m)!}}(-1)^me^{im\phi}P^m_l(\cos \theta)
%\ea
%where $P^m_l$ are the associated Legendre polynomials. Spherical harmonics with $m<0$ are defined in terms of those with $m >0$: 
%\be
%Y_{lm}=(-1)^m \overline{Y}_{l(-m)}
%\ee
%They are orthonormal (in $\Fl^2(S^2,\sin\theta\,d\phi d\theta)$)
%\ba
%\int_0^\pi \sin \theta \, d\theta \int_0^{2\pi} d\phi \overline{Y_{l'm'}(\theta,\phi)}Y_{lm}(\theta,\phi)=\delta_{l,l'}\delta_{m,m'}
%\ea
%and eigenfunctions of $L^2$ and $L_3$:
%\ba
%L_3 \,Y_{lm}&=&\hbar m\, Y_{lm} \\
%L^2 \,Y_{lm}&=&\hbar^2 l(l+1)\, Y_{lm}  \q.
%\ea

\subsection{Model with Structure Functions rather than Structure 
Constants Linear in the Momenta}
\label{s5.4}

As an example for a model with structure functions we will discuss a sort of a deformed $SO(3)$ constraint algebra
\ba
C_1=x_2 p_3^m -x_3^n p_2  \q \q\q
C_2=x_3^n p_1-x_1 p_3^m  \q \q\q
C_3=x_1 p_2-x_2 p_1   
\ea
\ba \label{SFcons1}
\{C_1,C_2\}=m\,n\, p_3^{m-1}x_3^{n-1} \,C_3 \q\q\q \{C_2,C_3\}=C_1 \q\q\q \{C_3,C_1\}=C_2  
\ea
where $n$ and $m$ are positive natural numbers. For $n=m=1$ we recover the $SO(3)$-algebra. One can do a similar deformation for $SO(p,q)$-algebras.

The Dirac observables (see \cite{PartObs1}) for this system generate an $sl(2,\Rl)$ algebra:
\ba
e^+&=&x_1^2+x_2^2+\frac{2}{m+n}p_3^{1-m}x_3^{n+1} \q\q \q\q 
e^-=p_1^2+p_2^2+\frac{2}{m+n}x_3^{1-n}p_3^{m+1} \nn \\ 
d&=&x_1p_1+x_2p_2+\frac{2}{m+n}x_3 p_3  
\ea
\ba
\{d,e^{\pm}\}&=&\mp 2e^\pm \q\q \q \{e^+,e^-\}=4d  \q.
\ea
For general $m,n$ these observables commute only on the constraint hypersurface with the constraints. 

We define the Master Constraint as
\be
\MC=C_1^2+C_2^2+C_3^2 \q.
\ee
 The constraints $C_1$ and $C_2$ do not strongly Poisson-commute with 
$\MC$ 
 (for $n \neq 1$ or $m \neq 1$), but $C_3$ does. For the rest of this 
 chapter, we will consider the case $m=1$ and $n$ odd. For these parameter 
 values the phase space function $e^+$ is non-negative and commutes on the 
 whole phase space $\Rl^3 \times \Rl^3$ with the Master Constraint. This 
 will be helpful for the spectral analysis of the quantized Master 
 Constraint. The case $n=1$ and $m$ odd can be treated in the same way, 
 using Fourier transformation. For $m>1$ the observable $e^+$ contains 
 negative powers of $p_3$, therefore the analysis gets more complicated. 

\subsubsection{Quantization}

We will quantize this system in the usual way by assigning multiplication operators $\hat x_i$ to the configuration variables $x_i$ and differential operators $\hat p_j=-i\hbar \partial_j$ to the momenta $p_j$.  The kinematical Hilbert space, we are starting with, is $L_2(\Rl^3)$.

 There arises no factor ordering ambiguity for the quantization of the 
 constraints, and since the structure function $f=n (\hat x_3)^{n-1}$ 
 commutes with $C_3$ (see (\ref{SFcons1})), it is possible to quantize all 
 the constraints as symmetric operators as we will explain in the 
conclusion section and to have $\hat C_3$ standing to 
 the right of the structure function: 
\ba
\big[\hat C_1, \hat C_2\big]=i\hbar\, n (\hat x_3)^n \,\hat C_3 \q\q 
\big[\hat C_2, \hat C_3 \big]=i\hbar \,\hat C_1 \q\q
\big[\hat C_3, \hat C_1 \big]=i\hbar \, \hat C_2 \q\q.
\ea

We have to analyze the Master Constraint Operator
\be
\MCO =\hat C_1^2+ \hat C_2^2 + \hat C_3^2 \q,
\ee
which is a second order partial differential operator. It can be densely defined and is symmetric on the linear span of the Hermite functions.

To solve the eigenvalue equation for the Master Constraint Operator we will introduce new coordinates $(t,\theta,\phi)$ analogous to the coordinates $(r,\theta,\phi)$ in the $SO(3)$ case. However, for $n \neq 1$ we have to consider the regions $x_3 \geq 0$ and $x_3 \leq 0$ seperately:
For $x_3 \geq 0$ we define
\ba
x_1 &=& t \, \cos \phi \, \sin \theta   \q\q t\geq 0\q \q \phi \in \big[ 0, 2\pi \big) \q\q \theta \in \big[ 0, \tfrac{\pi}{2}\big] \nn \\
x_2 &=& t\, \sin \phi \sin \theta \nn \\
x_3 &=& (\frac{n+1}{2})^{\frac{1}{n+1}} ( t\, \cos \theta)^{\frac{2}{n+1}} 
\ea
and for $x_3 \leq 0$
\be
x_3=-(\frac{n+1}{2})^{\frac{1}{n+1}} (- t\, \cos \theta)^{\frac{2}{n+1}} \q\q \theta \in \big[ \tfrac{\pi}{2}, \pi \big] \q.
\ee
The measure is transformed to $dx_1 dx_2 dx_3=  \sin \theta \,(\frac{2}{n+1})^\frac{n}{n+1} t^\frac{n+3}{n+1} |\cos \theta|^\frac{-n+1}{n+1} dt d\theta d\phi$.
With these coordinates we have
\be
x_1^2+x_2^2+\frac{2}{n+1} x_3^{n+1}=t^2=e^+
\ee
and $\MCO$ does not include derivatives with respect to $t$. Indeed, for $\cos\theta \geq 0$,
\ba
\MCO=-\hbar^2 \Big(\frac{n+1}{2}\Big)^\frac{2n}{n+1}\Bigg[(t \cos \theta)^\frac{2n-2}{n+1} \partial^2_\theta + \bigg( \frac{(t\cos\theta)^\frac{3n-1}{n+1}}{t \sin\theta}-t^\frac{2n-2}{n+1}(\cos\theta)^\frac{n-3}{n+1} \Big(\frac{n-1}{n+1}\Big) \sin\theta \bigg) \partial_\theta + \nn \\ \bigg( \frac{(t\cos\theta)^\frac{4n}{n+1}}{t^2 \sin^2 \theta}+\Big(\frac{2}{n+1}\Big)^\frac{2n}{n+1} \bigg)  \partial^2_\phi \Bigg] \q
\ea
and $\MCO$ for $\cos\theta \leq 0$ is obtained from the above formula by replacing $t$ with $-t$. The operator $\MCO$ simplifies considerably, if we introduce the coordinate
\ba
u&=&(\cos \theta)^\frac{2}{n+1}   \q \text{for}\q \cos\theta \geq 0 \nn \\
u&=&-(-\cos \theta)^\frac{2}{n+1}   \q \text{for}\q \cos\theta \leq 0 \q\q u \in [-1,1] \q.
\ea
The coordinate $u$ is proportional to the coordinate $x_3$ and therefore one gets the same operator for $x_3$ positive and for $x_3$ negative:
\ba
\MCO=-\hbar^2\Big(\frac{n+1}{2}\Big)^\frac{2n}{n+1}  \Bigg[\bigg(\frac{t^\frac{2n-2}{n+1}\,u^{2n}}{1-u^{n+1}}+\Big(\frac{2}{n+1}\Big)^\frac{2n}{n+1}\bigg)\partial^2_\phi+\frac{2t^\frac{2n-2}{n+1} }{n+1} \partial_u\bigg((1-u^{n+1})\partial_u\bigg)\Bigg]  
\ea
The measure is now $dx_1 dx_2 dx_3=(\frac{n+1}{2})^\frac{1}{n+1}t^\frac{n+3}{n+1}dt du d\phi$.

Since the operators $\hat C_3=-i\hbar\partial_\phi$ and $\MCO$ commute, we can diagonalize them simultaneously. Choosing periodic boundary conditions in $\phi$ one obtains $\text{spec}(\hat C_3)=\{\hbar k |k \in \Zl\}$ and the null eigenspace consists of constant functions in $\phi$. We are interested in the spectrum of $\MCO$ near zero, so it suffices to consider $\MCO$ restricted to the null eigenspace of $\hat C_3$, as it has elsewhere spectrum bounded from below by $\hbar ^2$ (since $\MCO=\hat C_1^2+\hat C_2^2+\hat C_3^2$). 
The restriction of $\MCO$ to this space is
\be
\MCO_{|\hat C_3=0}=-\hbar^2\Big(\frac{n+1}{2}\Big)^\frac{2n}{n+1}\frac{2t^\frac{2n-2}{n+1} }{n+1} \partial_u\bigg((1-u^{n+1})\partial_u\bigg) \q,
\ee
and can be seen as a product of two commuting operators $B_1$ and $B_2$:
\ba
B_1:=\hbar^2\Big(\frac{n+1}{2}\Big)^\frac{2n}{n+1}\frac{2t^\frac{2n-2}{n+1} }{n+1} \q,\q\q\q
B_2:=-\partial_u\bigg((1-u^{n+1})\partial_u\bigg) \q.
\ea

In the following, we will consider the operator $B_2$ on the Hilbert space $L_2((-1,1),du)$ and show that its spectrum is purely discrete. 
Afterwards we will come back to the product of the operators $B_1 \cdot B_2$. 
(The product $B_1 \cdot B_2$ cannot be seen as a tensor product of $B_1$ and $B_2$ since $L_2(\Rl^3, dx^3)$ is not equivalent to $L_2(\Rl_+,t^\frac{n+3}{n+1}dt)\otimes L_2((-1,1),du)\otimes L_2((0,2\pi),d\phi)$. 
Moreover one has to be careful, since it is not guaranteed that $B_2$ has a self-adjoint extension as an operator on $L_2(\Rl_3)$. 
This might be the case, because $B_2$ does not preserve the space of Hermite functions in the three variables $(x_1,x_2,x_3)$, only the combination $B_1\cdot B_2$ does preserve this space. Nevertheless one can consider the operator $B_2$ defined on the Hilbert space $L_2((-1,1),du)$.)

%The operator $B_1$ is a (positive) multiplication operator, therefore its domain of selfadjointness is
%\be
%\Fd_{SA}(B_1)&=&\{f \in  L_2(\Rl_+,t^\frac{n+3}{n+1}dt)|B_1\cdot f \in L_2(\Rl_+,t^\frac{n+3}{n+1}dt) \} \q.
%\ee

The operator $B_2$ is symmetric and positive on the dense domain
\ba \label{domainB2}
\Fd(B_2)=\{f \in  L_2((-1,1),du)|f \in AC(-1,1), (1-u^{n+1})f' \in AC(-1,1), \nn \\
 B_2\cdot f \in L_2((-1,1),du) ,\underset{u\rightarrow\pm 1}{\text{lim}}(1-u^{n+1})f'(u)=0\} 
 \ea
where $AC$ denotes the class of absolutely continuous functions.
The positivity can be seen by using integration by parts
\be
<f,B_2 \cdot f>_{L_2(u)}=\int_{-1}^1 \overline{\partial_u f(u)}(1-u^{n+1})\partial_u f(u) \,du \,\,\geq 0
\ee
for functions $f \in \Fd(B_2)$. A positive and symmetric operator has 
always selfadjoint extensions (see \cite{ReedSimonII}).

 The operator $B_2$ is an ordinary differential operator, more specifically it is a Sturm-Liouville operator on the interval $I:=(-1,1)$, so we can utilize the theory of Sturm-Liouville operators, see for example \cite{SchroedOp}. Sturm-Liouville operators can be classified according to the behaviour of eigensolutions at the endpoints of the interval $I$: A Sturm-Liouville operator $A$ is limit circle at the endpoint $a$ if for one $\lambda \in \Cl$ all solutions to $(A-\lambda)f=0$ are square integrable near $a$. One can prove, that if this is the case for one $\lambda$, then it holds for all $\lambda \in \Cl$. It therefore suffices to consider the solutions to $B_2\cdot f=\lambda f$ for one $\lambda \in \Cl$, in particular $\lambda=0$. 

One solution for $\lambda=0$ is obviously given by $f_{01}(u) \equiv 1$, the 
other solution is given by
\be
f_{02}(u)=\int_0^u \frac{1}{(1-u'^{n+1})(f_{01}(u'))^2}du'=\int_0^u \frac{1}{(1-u'^{n+1})}du' \q.
\ee
Both solutions are square integrable on $(-1,1)$. The function $f_{02}$ is 
integrable because for $n>1$ and $n$ odd, the function $|f_{02}(n>1)|$ is 
less than $|f_{02}(n=1)|$ on the interval $(-1,1)$. But $f_{02}(n=1)=\text{artanh}(u)$ is square integrable on $(-1,1)$. We are therefore in the limit circle case for both boundaries $u_\pm=\pm1$.

% limit circle: both solutions are square integrable, deficiency index is (1,1) or (2,2): one has to define boundary conditions (possible bdry cond. constitute a circle)
% limit point: deficiency index is (0,0), op. is ess. s.a.;  particle stays away from the bdr.ies, so that it is not neseccary to define bdry cond.

 If both boundary points are limit circle for an operator $A$, the 
 following holds (see \cite{SchroedOp}): The spectrum of any self adjoint 
 extension is purely discrete, the eigenfunctions are simple 
(multiplicity one) and form an 
orthonormal basis and the resolvent of $A$ is a Hilbert -- Schmidt 
operator. Notice that Hilbert -- Schmidt operators are in particular 
compact, hence the spectrum of the resolvent $R_z(A)$ for $z\in\Rl$ not 
in the spectrum of $A$ has an accumulation point at most at zero. Thus 
$A$ does not have any accumulation point and since $\MCO$ is unbounded 
we know that the eigenvalues actually diverge (when ordered according to 
size).

Hence $B_2$ has a discrete simple spectrum, and since $f_{01}(u) \equiv 1$ is 
a square integrable solution for $\lambda=0$ 
(which fulfills contrary to $f_{02}$ the boundary conditions), zero is 
included in the spectrum. 
This holds for any selfadjoint extension of $B_2$. Now $B_2$ with the 
domain (\ref{domainB2}) is positive. 
We choose in the following any selfadjoint extension
such that it is still positive and such that $B_1\cdot B_2$ as an operator 
in 
$L_2(\Rl^3)$ corresponds to the positive self-adjoint extension of $\MCO$, 
we have chosen before (e.g. the Friedrich extension)\footnote{The 
operator $B_1$ is self-adjoint with 
the domain $\Fd_{SA}(B_1)=\{f \in  L_2(\Rl^3)|B_1\cdot f \in L_2(\Rl^3) 
\}$.}.
%(This can always be done, for instance via the Friedrich's extension, see \cite{ReedSimonII}.) 
The eigenvalues of $B_2$ are therefore positive 
(that is $\lambda_j \geq 0$ and $\lambda_j=0$ if and oly if 
$j=0$).

Since the operator $B_2$ has a null eigenspace consisting of functions constant in $u$ and therefore $\theta$, one would expect that the physical Hilbert space consists of functions functionally independent of $\theta$ (and $\phi$ as was argued above). However we are dealing not just with the operator $B_2$ but with the product $B_1 \cdot B_2$. The operator $B_1$ is a multiplication operator and has continuous spectrum on the whole positive axis including zero. 

Therefore we have to discuss the decomposition of the Hilbert space into a direct integral Hilbert space and to calculate the spectral measures. To this end we will consider the restriction of the Master Constraint Operator to the subspace $\hat{C_3}=0$ as it has elsewhere a spectrum, which is bounded from below by $\hbar^2$. Since this subspace coincides with functions in $L_2(\Rl^3)$, which are constant in the angle variable $\phi$, we can identify this subspace with $\ch':=L_2(\Rl_+ \times \Rl, \rho\, d\rho dx_3)$ where the variable $\rho$ is defined by $\rho^2=x_1^2+x_2^2$. 

A dense system of vectors in this Hilbert space is generated by the set
\be \label{strfctsdense}
\{f_{pq}:=\rho^{2q} x_3^p\, \exp(-\tfrac{1}{2}t^2) \q|\q q,p \in \Nl \}  \q .
\ee
We need just the even polynomials in $\rho$ since the variable $\rho$ 
has range only in the positive half axis\footnote{Indeed, polynomials
of any positive power of $t^r$ could be used in order to construct a basis
of the Hilbert space by the Gram -- Schmidt procedure because any 
square integrable function $t\mapsto f(t)$ can be written as 
$t\mapsto f_r(t^r)$ where $f_r(s):=f(s^{1/r})$. Notice that we do not 
transform the measure to the variable $t^r$ here.} $\Rl_+$.
Now one has to find a cyclic 
system of vectors with respect to $\MCO$. We will begin with functions that are polynomials in $t^2=(\rho^2+\tfrac{2}{n+1} x_3^{n+1})$ weighted with $\exp(-t^2)$. All these functions are annihilated by $\MCO$, hence to get a cyclic system, we need all the even powers of $t$. (Again, we just need the even powers, since $t$ extends over the positive half axis.) If one performs the Gram-Schmidt procedure for this system $\{t^{2k} \exp(-t^2)\,|\, k\in \Nl \}$ one will get an ortho-normal set of vectors $\{v_{0k}\,|\, k\in \Nl \}$ which can be identified as a basis in the Hilbert space $L_2(\Rl_+,t^\frac{n+3}{n+1}dt)$. The ascociated spectral measures are given by     
\ba
\mu_{0k}(\lambda)=<v_{0k},\Theta(\lambda-B_1\cdot B_2)v_{0k}>=<v_{0k},\Theta(\lambda-B_1\cdot 0)v_{0k}>=\Theta(\lambda)\q ,
\ea
so that the set of vectors $\{v_{0k}\}$ is associated with a pure point spectral measure. We will call the subspace spanned by these vectors $\ch_{pp}$ (anticipating that all vectors orthogonal to this subspace are associated to spectral measures, which are absolutely continuous).

Now we have to find the orthogonal complement to $\ch_{pp}$. If one rewrites 
the functions $f_{pq}$ in terms of the coordinates $(t,u)$ one gets
\ba \label{0.0}
 f_{pq}=\rho^{2q} x_3^p=C(p,q) (1-u^{n+1})^q u^p t^{2q+\frac{2p}{n+1}}  e^{-\tfrac{1}{2} t^2}
\ea
where $C(p,q)$ is a $p,q$ dependent constant. 
It is important to note here that a power of $u^a$ is alway accompanied by a power of at least $t^\frac{2a}{n+1}$. 
The underlying reason for this is that $u\sim x_3t^{-\frac{2}{n+1}} $. 
Using this fact, it is straightforward to see, that also 
\ba
g_{pq}=u^p t^{\frac{2p}{n+1}+2q}e^{-\tfrac{1}{2} t^2} \q p,q \in \Nl
\ea
or
\ba
h_{pq}=P_p(u) t^{\frac{2p}{n+1}+2q}    e^{-\tfrac{1}{2} t^2}
\ea
generate a dense subspace in $\ch'$. 
Here $P_p(u)$ are the Legendre polynomials of order $p$, 
that is $P_p(u)$ is a polynomial with degree $p$. 
To see this, notice that  (\ref{0.0})
is a polynomial in terms of $t^2$ and $x:=ut^{2/(n+1)}$ which explains the 
statement for $g_{pq}$. Next notice that $P_p$ is a polynomial of order 
$p$ in $u$ which can be written as a polynomial in $x$ of the same 
order and 
$t^{2(p-k)/(n+1)},\;k=0,..,p$. Now any positive power of $t$ can be 
expanded in terms of the $v_{0k}$ and thus in terms of the $t^{2q}$ again.  

Since $P_0(u)$ is the constant function, the space spanned by $\{h_{0q}\,|q\in \Nl\}$ coincides with the space spanned by $\{v_{0k}\,|k\in \Nl\}$. 
Moreover $P_p,p>0$ is orthogonal to $P_0$ (in $L_2((-1,1),du)$), so that the orthogonal complement to $\ch_{pp}$ is spanned by $\{h_{pq}\,|\Nl \ni p>0,q\in \Nl\}$. 
We will now show, that the spectral measures associated to vectors from this subspace are absolutely continuous. In order to do this we have to expand the functions $P_p(u) \in L_2((-1,1),du),p>0$ into eigenfunctions of $B_2$:
\ba
P_p(u)=\sum_{k=1}^\infty a^p_k \psi_k(u)
\ea
where $\psi_k$ is the normalized $k-th$ eigenfunction of $B_2$ (and we assume that the corresponding eigenvalues $\lambda_k$ are ordered and that $\psi_0$ is the constant function).
 The constant function (i.e. $\psi_0$) does not appear in this decomposition since $P_p$ is orthogonal (in $L_2((-1,1),du)$) to the constant function for $p>0$.

Using this decomposition and abbreviating $c_n=\hbar^2(\frac{n+1}{2})^\frac{2n}{n+1}\frac{2}{n+1}$ we can write
\ba
\mu_{p'q'pq}(\lambda)&:=&<h_{p'q'},\theta(\lambda-\MCO)h_{pq}>  \nn \\
&=& \Big(\frac{n+1}{2}\Big)^\frac{1}{n+1} 
\int_{\Rl_+}\int_{[-1,1]} dtdu \,
\exp(-t^2)\,  t^{\frac{2(p+p')}{n+1}+2(q+q')+\frac{n+3}{n+1}}
\Bigg(\sum_{k'=1}^\infty \overline{a}^{p'}_{k'} \overline{\psi}_{k'}(u)\Bigg) \nn\\
&& \q\q\q\q\q\q\q\q\q\q
\theta(\lambda-c_n t^\frac{2n-2}{n+1} B_2)
\Bigg(\sum_{k=1}^\infty a^{p}_{k} \psi_{k}(u)\Bigg) \nn \\
&=&
 \Big(\frac{n+1}{2}\Big)^\frac{1}{n+1}   \sum_{k',k=1}^\infty 
\int_{\Rl_+}\int_{[-1,1]} dtdu \,
\exp(-t^2)\,  t^{\frac{2(p+p')}{n+1}+2(q+q')+\frac{n+3}{n+1}} \nn \\
&& \q\q\q\q\q\q\q\q\q\q
\theta(\lambda-c_n t^\frac{2n-2}{n+1} \lambda_k)\,\,\overline{a}^{p'}_{k'}a^{p}_{k}\,\,
\overline{\psi}_{k'}(u)\psi_{k}(u) \nn \\
&=&
\Big(\frac{n+1}{2}\Big)^\frac{1}{n+1}   \sum_{k=1}^\infty \int_{\Rl_+}dt\,
\exp(-t^2)\,  t^{\frac{2(p+p')}{n+1}+2(q+q')+\frac{n+3}{n+1}} \nn \\
 && \q\q\q\q\q\q\q\q\q\q
\theta(\lambda-c_n t^\frac{2n-2}{n+1} \lambda_k)\,\,\overline{a}^{p'}_{k}a^{p}_{k}
\ea
where in the last line we used the ortho-normality of $\psi_k$ in $L_2((-1,1),du)$. 

The sum in the last line converges absolutely, since the integral over $t$ can be bounded ($k$-independently) from above by  ignoring the $\theta$-function. 
Then we are left with the sum over the absolute values of $\overline{a}^{p'}_{k}a^{p}_{k}$, which can be estimated using the Cauchy-Schwarz inequality and the fact that the $a^p_k$ are the expansion coefficients of $P_k \in L_2((-1,1),du)$.

 This shows that all the measures $\mu_{p'q'pq}$ are absolutely continuous 
with respect to the Lebesgue measure on $\Rl_+$. 
Indeed we have
\ba
\frac {d\mu_{p'q'pq}(\lambda)}{d\lambda}
&=&
\Big(\frac{n+1}{2}\Big)^\frac{1}{n+1}   \sum_{k=1}^\infty \overline{a}^{p'}_{k}a^{p}_{k} \int_{\Rl_+}dt\,
\exp(-t^2)\,  t^{\frac{2(p+p')}{n+1}+2(q+q')+\frac{n+3}{n+1}} \delta(\lambda-c_n t^\frac{2n-2}{n+1} \lambda_k) \nn \\
&=& 
\Big(\frac{n+1}{2}\Big)^\frac{1}{n+1} \sum_{k=1}^\infty \overline{a}^{p'}_{k}a^{p}_{k} \,
\big(\frac{n+1}{2c_n\lambda_k(n-1)} \big)
\exp\!\big(-(\frac{\lambda}{c_n\lambda_k})^\frac{n+1}{n-1}\big)
 \,\big(\frac{\lambda}{c_n\lambda_k}\big)^\frac{3+p+p'+(q+q')(n+1)}{n-1} \nn \\
&=:&
 \lambda^\frac{3+p+p'+(q+q')(n+1)}{n-1} f_{p'q'pq}(\lambda)
\ea
where in the second line we performed a coordinate transformation 
$s_k=c_n \lambda_k t^\frac{2n-2}{n+1}$ in order to solve the 
delta-function. The sum in the second line converges absolutely 
(uniformly in $\lambda$) since the exponential factor can be estimated by 
 $1$ and the $\lambda_k$ have to be bigger than $1$ for some finite $k$ 
% (otherwise $B_2$ would be bounded, which is not the case)
because the spectrum of $B_2$ does not have an accumation point as we 
showed above.
Hence 
 $f_{p'q'pq}$ is well defined, in particular $f_{p'q'pq}(0) >0$ for $p'=p$ 
(since all terms in the sum are then positive).

To summarize, our Hilbert space $\ch'$ can be decomposed into two subspaces 
$\ch_{pp}$ and $\ch_{ac}$ where $\MCO$ restricted to $\ch_{pp}$ has a pure 
point spectral measure and $\MCO$ restricted to $\ch_{ac}$ has an absolutely 
continuous spectral measure. 
%From the discussion we know that in 
%the direct integral decomposition the contribution of $\ch_{ac}$ is 
%surpressed on the support of the pure point measure.
 We therefore have to discuss the direct integral decomposition of 
 $\ch_{pp}$ and $\ch_{ac}$ separately. The `direct integral decomposition' 
 for $\ch_{pp}$ is straightforward: Since $\ch_{pp}$ is a proper 
eigenspace 
 to to the null eigenvalue of $\MCO$, the physical Hilbert space 
 corresponding to the pure point spectrum coincides with $\ch_{pp}$ (which 
can be identified with $L_2(\Rl_+,t^\frac{n+3}{n+1}dt)$).

%The decomposition of $\ch_{pp}$ is easy, since the spectral measures $\mu_{0k}$ are all the same and the support of these measures is just $\{0\}$. The induced Hilbert space $\ch_{pp,ind}$ is therefore the whole space $\ch_{pp}$, which can be identified with the space of functions constant in $\theta$ and $\phi$, i.e $L_2(\Rl_+,t^\frac{n+3}{n+1} dt)$. This is exactly the result one would expect beforehand.

To perform the direct integral Hilbert space decomposition of $\ch_{ac}$ spanned by $\{h_{pq}\,|\Nl\ni p>0,q\in \Nl\}$ we choose an ortho-normal cyclic system $\{\Omega_m\, | \Nl \ni m >0\}$ in that space, such that $\Omega_1=N_1 h_{10}$, where $N_1$ is a normalization constant. Since $\{h_{pq}\,|\Nl\ni p>0,q\in \Nl\}$ generates $\ch_{ac}$ we can always find coefficients $A^m_{pq}$ such that
\ba
\Omega_m=\sum_{p=1,q=0}^\infty A^m_{pq} h_{pq}   \q  .
\ea

%\vspace{1cm}
%{\it It would be nice if one could show, that the sum includes only 
%finitely many terms!!!!! But that seems to be difficult because of the 
%action of $\MCO$. Some comments on convergents of the stuff below?? (But 
%one can  use that $<\Omega_k,\Omega_k>=1$. Expand that into $h_{pq}$ 
%Moreover for $\lambda$ near zero $\mu_{p'q'pq}$ are very small and zero 
%for $\lambda=0$.) }
%\vspace{1cm}

Consider the spectral measures $\mu_m,m>1$ associated to this cyclic system
\ba
\mu_m(\lambda)&:=&<\Omega_m,\theta(\lambda-\MCO)\Omega_m> \nn \\
       &=& <\Omega_m,\, \theta(\lambda-\MCO) \sum_{p=1,q=0}^\infty A^m_{pq} h_{pq}> \nn \\
&=& <\Omega_m,\, \theta(\lambda-\MCO) \sum_{\stackrel{p=1,q=0}{(p,q)\neq (1,0)}}^\infty A^m_{pq} h_{pq}>
\ea
since $<\Omega_m, \theta(\lambda-\MCO)\Omega_1>=N_1<\Omega_m, \theta(\lambda-\MCO)h_{10}>=0$ due to the defining property of a cyclic system. Hence
\ba
\mu_m(\lambda)&=&\sum_{p'=1,q'=0}^\infty\sum_{\stackrel{p=1,q=0}{(p,q)\neq (1,0)}}^\infty \overline{A}^m_{p'q'} A^m_{pq} <h_{p'q'},\theta(\lambda-\MCO) h_{pq}> \nn \\
&=& \sum_{p'=1,q'=0}^\infty\sum_{\stackrel{p=1,q=0}{(p,q)\neq (1,0)}}^\infty \overline{A}^m_{p'q'} A^m_{pq}\,\,\, \mu_{p'q'pq}(\lambda)  \q  .
\ea
Differentiating under the sums\footnote{Strictly speaking one must verify
that we may interchange summation and differentiation. This could be done 
for instance by verifying that the series for $\mu_m(\lambda)$ 
converges absolutely and uniformly in $\lambda$ at least close to zero 
(the absolute 
and uniform convergence of the series for $\mu(\lambda)$ then follows).
We have not checked that this is the case but given the fact that the 
$\mu_{p'q'pq}$ 
converge rapidly to zero in the vicinity of $\lambda=0$ this should 
be true. With the tools given here, one could check this explicitly
by a tedious but straightforward calculation.}
\ba
\frac{d\mu_m(\lambda)}{d\lambda}
&=& \sum_{p'=1,q'=0}^\infty\sum_{\stackrel{p=1,q=0}{(p,q)\neq (1,0)}}^\infty \overline{A}^m_{p'q'} A^m_{pq}\,\,\, \frac{d\mu_{p'q'pq}(\lambda)}{d\lambda} \nn \\
&=&  \lambda^\frac{6}{n-1}\sum_{p'=1,q'=0}^\infty\sum_{\stackrel{p=1,q=0}{(p,q)\neq (1,0)}}^\infty \overline{A}^m_{p'q'} A^m_{pq}\,\,\, 
\lambda^\frac{p+p'+(q+q')(n+1)-3}{n-1} f_{p'q'pq}(\lambda) \nn \\
&=:&
\lambda^\frac{6}{n-1} g_m(\lambda)
\ea
 where $g_m(\lambda)$ is a non-singular function of $\lambda$ near zero 
 (that is $g_m(0)$ is either zero or some finite value). This holds for 
 $m>1$; for $m=1$ we get
\ba
\frac{d\mu_1(\lambda)}{d\lambda}=\lambda^\frac{5}{n-1} (N_1)^2f_{1010}(\lambda)
\ea
where $f_{1010}(0)=c>0$.

The total spectral measure $\mu_{ac}(\lambda)$ on $\ch_{ac}$ is given 
by (if the maximal multiplicity is less than infinity, choose appropriate 
normalization constants different from $2^{-m}$)
\ba
\frac{d\mu_{ac}(\lambda)}{d\lambda}&:=&\frac{1}{2}\frac{d\mu_1(\lambda)}{d\lambda}+\sum_{m=2}^\infty 2^{-m} \frac{d\mu_m(\lambda)}{d\lambda} \nn \\
&=& 
\frac{1}{2}\lambda^\frac{5}{n-1} (N_1)^2f_{1010}(\lambda)+\lambda^\frac{6}{n-1}\sum_{m=2} 2^{-m} g_m(\lambda)     \q .
\ea
Now we can consider the Radon-Nikodym derivatives in the limit $\lambda\rightarrow 0$. Assume that $m>1$:
\ba
\lim_{\lambda\rightarrow 0} \rho_m(\lambda) 
=
\lim_{\lambda\rightarrow 0} \frac{d\mu_m(\lambda)/d\lambda}{d\mu_{ac}(\lambda)/d\lambda} =\lim_{\lambda\rightarrow 0} \frac{\lambda^\frac{6}{n-1} g_m(\lambda)}{ \frac{1}{2}\lambda^\frac{5}{n-1} (N_1)^2f_{1010}(\lambda)+\lambda^\frac{6}{n-1}\sum_{m=2} 2^{-m} g_m(\lambda) }
=0\q
\ea
Hence all the $\rho_m,m>1$ vanish at $\lambda=0$. But the same calculation for $m=0$ gives $\rho_1(0)=2$. 
Therefore the induced Hilbert space for $\lambda=0$ from $\ch_{ac}$ is one-dimensional, that is unitarily equivalent to $\Cl$. Putting the contributions from $\ch_{pp}$ and $\ch_{ac}$ together we get
$
\ch_{phys}\simeq L_2(\Rl_+, t^\frac{n+3}{n+1}) \oplus \Cl
$
(where the inner product in $\Cl$ can be rescaled arbitrarily). The first 
term corresponds to the (proper) null subspace with respect to $\MCO$ the 
second to the continious spectrum of the multiplication operator in $t$.\\
\\
Remark:\\
By the standard theory for Sturm -- Liouville operators such as $B_2$ 
we know that the eigenfunction $\psi_k$ has $k$ zeroes and that the 
eigenvalues asymptote to $\lambda_k\propto k^2$. By the Weierstrass 
theorem, $\psi_k$ can be approximated in the sup -- norm arbitrarily well 
on the compact set $[-1,1]$ by polynomials of degree at least $k$ (in 
order to have $k$ real roots). Thus, the $\psi_k$ are actually not too 
different from the standard Legendre polynomials $P_k$ and if they would 
be really polynomials of degree $k$ we could just use the functions 
$\Omega_k=\psi_k t^{2k/(n+1)}$ as a cyclic system. Then the above 
calculations would become entirely trivial because the Radon -- Nikodym 
derivative at $\lambda=0$ would be obviously non -- vanishing for the 
lowest order $k=1$. Unfortunately the eigenvectors $\psi_k$ are 
not polynomials unless $n=1$ and so we had to go through this 
very elaborous analysis.

\section{Pure Point and Absolutely Continuous Spectra} 
\label{pp and ac}

 Here we will discuss an example, where, similarly to the previous one, 
the Master Constraint Operator has 
 pure point and absolutely continuous spectrum at zero. The example is 
simpler without structure constants and serves the purpose to illustrate 
an important point when choosing the cyclic system.

 We will start with a kinematical Hilbert space 
$L_2(\Rl^2)$ and a Master Constraint Operator
\be
\MCO=(\hat{x}_1^2+\hat{x}_2^2)(\hat{x}_1\hat{p}_2-\hat{x}_2 \hat{p_1})^2
\ee
 where the $\hat{x}_i$ are multiplication operators and the 
 $\hat{p}_i:=-i\hbar \partial_i$ act by differentiation. Apriori one would 
 expect that the physical Hilbert space includes the space of functions 
 with zero angular momentum and a one-dimensional part which corresponds 
 to the generalized eigenfunction $\delta(x_1)\cdot\delta(x_2)$. 

We will now construct the direct integral decomposition of the kinematical Hilbert space with respect to $\MCO$. First of all, we have to find a cyclic basis for $L_2(\Rl^2)$. To this end we note, that $L_2(\Rl^2)$ is spanned by polynomials in $x_1,x_2$ weighted by a gaussian factor. These can be generated by the set
\be
\{ x_1^{n_1} x_2^{n_2} \exp(-\tfrac{1}{2}(x_1^2+x_2^2)\,|\, n_1,n_2 \in \Nl\}
\ee
and also by the set
\be \label{vmn}
\{ v_{mn}=x_+^m x_-^n\,\exp(-\tfrac{1}{2}x_+x_-)\,|\, n,m \in \Nl\}
\ee
where we defined $x_\pm:=x_1\pm i x_2$. In spherical coordinates defined by
\ba
x_1&=& r\,\cos\phi \q\q   r\in \Rl_+, \phi \in [0, 2\pi) \nn \\
x_2&=& r\, \sin\phi
\ea
the vectors $v_{nm}$ can be expressed as
\be
v_{nm}= r^{n+m}\exp(i\phi(m-n))\; e^{-r^2/2} \q .
\ee
By introducing new indices $N:=(n-m)\in \Zl$ and $k:=\tfrac{1}{2}(m+n-|n-m|)\in \Nl$ the set (\ref{vmn}) can also be written as
\be
\{v_{Nk}=r^{|N|+2k} \exp(iN\phi)\exp(\tfrac{1}{2}r^2) \,|\, N\in \Zl, k\in \Nl\}\q.
\ee
 Applying repeatedly $\MCO$ to a vector $v_{N0}, N\neq 0$ generates all 
 the other vectors $v_{Nk},k\in \Nl$ with the same index $N$. We therefore 
 need just the $v_{N0}$ as cyclic vectors. But for $N=0$ applying $\MCO$ 
 to $v_{0k}$ gives zero, so that we also need the whole set $\{v_{0k}\,| 
 \,k\in \Nl\}$. However, notice that this set is not orthonormal, one has 
 therefore to perform a Gram-Schmidt procedure. This will result in an 
 orthonormal set of the form $\{\Omega_{0k}:=f_k(r^2)\exp(-r^2/2)\}$ where 
 the $f_k$ are polynomials of order $k$ in $r^2$. This set can be taken
 as a basis for $L_2(\Rl_+,rdr)$.  \\
\\
Remark:\\
At this point it is important to draw attention to the following subtlety:
Recall the definition of the tensor product of two Hilbert spaces 
${\cal H}_j$: This is the Hilbert space ${\cal H}_1\otimes {\cal H}_2$
consisting of pairs $\psi_1\otimes \psi_2:=(\psi_1,\psi_2)$ with 
$\psi_j\in {\cal H}_j$ equipped with the inner product 
$<\psi_1\otimes \psi_2,\psi'_1\otimes \psi'_2>_{{\cal H}_1\otimes{\cal 
H}_2}:=\prod_{j=1}^2 <\psi_j,\psi'_j>_{{\cal H}_j}$. One can show that
if $b^{(j)}_k$ is a basis for ${\cal H}_j$ then 
$b^{(1)}_k\otimes b^{(2)}_n$ is a basis for ${\cal H}_1\otimes {\cal 
H}_2$. If $(X_j,{\cal B}_j,\mu_j)$
are measure spaces and ${\cal H}_j:=L_2(X_j,d\mu_j)$ are separable Hilbert 
spaces then one can show by using Fubinis theorem that ${\cal H}_1\otimes 
{\cal H}_2$ is isometrically isomorphic 
to $L_2(M_1\times M_2,d\mu_1\otimes d\mu_2)$ via 
$\psi_1\otimes\psi_2\mapsto \psi_1(x_1)\psi_2(x_2)$ where 
$\mu_1\otimes \mu_2$ is the measure on $M_1\times M_2$ based on the 
smallest $\sigma-$algebra ${\cal B}_1 \otimes {\cal B}_2$ containing the 
``rectangles'' $B_1\times B_2$ where $B_j \in {\cal B}_j$ and by 
definition $\mu_1\otimes \mu_2(B_1\times B_2):=\mu_1(B_1)\mu_2(B_2)$.

Now consider the problem at hand:
In polar coordinates we have $L_2(\Rl^2,d^2 x)=L_2(\Rl^2,rdr d\phi)$ with 
$\Rl^2=[(\Rl_+-\{0\})\times S^1]\cup\{(0,0)\}$. One could now think that 
since the one point sets $\{(0,0)\}$ and $\{0\}$ respectively have $d^2x$ 
and $dx$ Lebesgue measure zero respectively that
\ba \label{0.1}
L_2(\Rl^2,d^2x)
&=& L_2(\Rl^2-\{(0,0)\},d^2 x)=
L_2((\Rl_+-\{0\})\times S^1,d^2x)
\nonumber\\
&=& L_2((\Rl_+ - \{0\},r dr)\otimes L_2(S^1,d\phi)
=L_2((\Rl_+,r dr)\otimes L_2(S^1,d\phi)
\ea
If (\ref{0.1}) would be true then, given some ONB $b{(1)}_k(r)$ for 
$L_2(\Rl_+,r dr)$ consisting of polynomials in $r$ times 
$e^{-r^2/2}$ obtained via the 
Gram -- Schmidt procedure and an ONB 
$b^{(2)}_n(\phi)=e^{in\phi}/\sqrt{2\pi}$ for 
$L_2(S^1,d\phi)$ we would obtain a basis $b^{(1)}_k\otimes b^{(2)}_n$ for 
$L_2(\Rl^2,d^2x)$. However, in the tensor product we then 
obtain a dense set of vectors of the form $r^k e^{in\phi} e^{-r^2/2}$ 
where the pair $(k,n)\in \Nl_0\times \Zl$ is unrestricted. On the other 
hand 
the Hermite polynomial basis for $L_2(\Rl^2,d^2x)$ provide a dense set
consisting of vectors of the form $r^{|n|+2l} e^{in\phi} e^{r^2/2}$ as we 
have just 
seen. We conclude that the basis $b^{(1)}_k\otimes b^{(2)}_l$ for 
$L_2((\Rl_+,r dr)\otimes L_2(S^1,d\phi)$ is overcomplete for 
$L_2(\Rl^2,d^2x)$ and hence these Hilbert spaces are not isometrically
isomorphic. In other words, the functions $r^k e^{in\phi} e^{-r^2/2}$ such 
that
$k-|n|$ is not a non -- negative and even integer could be expanded in 
terms of Hermite polynomials because they are obviously square integrable.

What has gone wrong in (\ref{0.1}) is that on $\Rl^2$ the coordinates 
$(r,\phi)$ are singular at $r=0$. The coordinates $x_1,x_2$ are globally 
defined which results in the fact that there is the restriction
$k-|n|=2l,\;l=0,1,2,..$ in order that the functions $r^k e^{in\phi}$ are 
are regular at $r=0$ when expressed in terms of $x_1,x_2$. This 
topological subtlety that $\Rl^2$ is a plane while $\Rl_+\times S^1$ 
is a half -- infinite cylinder with different orthonormal bases is of 
outmost importance for the Direct Integral Decomposition (DID) because 
neglecting this difference would result in a much bigger physical Hilbert 
space as we will see shortly.\\
\\
The normalization of the vectors $v_{N0}$ gives
\be
\Omega_{N0}=(\pi\,|N!|)^{-1/2} r^{|N|}\exp(iN\phi)\exp(-r^2/2) \q .
\ee
Their associated spectral measure are
\ba
\mu_{N0}=<\Omega_{N0},\Theta(\lambda-\MCO)\Omega_{N0}>&=&\frac{2}{|N|!}\int_0^\infty r^{2|N|} \exp(-r^2)\,\Theta(\lambda-r^2 N^2)\, rdr  \nn \\
&=& \frac{1}{|N|!}\int_0^\infty x^{|N|}\exp(-x) \, \Theta(\lambda-N^2 x)\,dx
\ea
which shows, that these measures are absolutely continuous with respect to the Lebesgue measure on $\Rl_+$.

The spectral measures associated to the $\Omega_{0k}$ can be calculated to
\be
\mu_{0k}=<\Omega_{0k},\Theta(\lambda-\MCO)\Omega_{0k}>=\Theta(\lambda)<\Omega_{0k},\Omega_{0k}>=\Theta(\lambda)
\ee
which shows, that these measures are of pure point type.

Hence the kinematical Hilbert space $L_2(\Rl^2)$ decomposes into a direct sum of two Hilbert spaces $\ch_{ac}$ and $\ch_{pp}$, on which $\MCO$ has either absolutely continuous or pure point spectrum. $\ch_{ac}$ is defined to be the closure of the span of $\{\MCO^k \Omega_{N0}\, |\, N\neq 0,N\in \Zl, k\in \Nl\}$ and $\ch_{pp}$ is the closure of the span of $\{\Omega_{0k}\,|\,k\in \Nl\}$. We will now discuss the direct integral decomposition on each of these two spaces seperately.

The direct integral decomposition of $\ch_{pp}$ (which can be identified with $L_2(\Rl_+, rdr)$ is easy to obtain, since all the measures $\mu_{0k}$ are the same and have just the point zero in their support. Hence $\ch_{pp}$ is alredy decomposed with respect to $\MCO$ -- the null eigenspace of $\MCO$ on $\ch_{pp}$ coincides with $\ch_{pp}$. The contribution from the pure point part of the spectrum to the physical Hilbert space is therefore given by $\ch_{pp}$. 

 For the direct integral decomposition of $\ch_{ac}$ we have to  
calculate first the total spectral measure 
\ba
 \mu_{ac}(\lambda)&=&\tfrac{1}{2}\sum_{N=1}^\infty 2^{-N}\left( 
\mu_{N0}(\lambda)+\mu_{-N0}(\lambda)\right) \nn\\
&=& \int_0^\infty \left( \sum_{N=1}^\infty \frac{(x/2)^N}{N!} \, \Theta(\lambda-N^2\, x) \right) \exp(-x) \, dx \q .
\ea
This gives for the Radon-Nikodym derivative with respect to the Lebesgue 
measure
\be
\frac{d\mu_{ac}}{d\lambda}= \sum_{N=1}^\infty \frac{1}{N^2\,N!} 
\left(\frac{\lambda}{2 N^2}\right)^N \exp(-\lambda/N^2) \q .
\ee
The Radon-Nikodym derivative of the measures $\mu_{N0}$ with respect to the Lebesgue measure is
\be
\frac{d\mu_{N0}}{d\lambda}=\frac{\lambda^{|N|}}{N^2\, 
|N|!}\exp(-\lambda/N^2)
\ee
so that the Radon-Nikodym derivatives of $\mu_{N0}$ with respect to $\mu_{ac}$ can be calculated to
\be
\rho_{N0}(\lambda)=
\frac{d\mu_{N0}/d\lambda}{d\mu_{ac}/d\lambda}=\lambda^{|N|}\frac{(N^2\, 
|N|!)^{-1}\exp(-\lambda/N^2)}{\sum_{M=1}^\infty \frac{1}{M^2\,M!} 
\left(\frac{\lambda}{2 M^2}\right)^M \exp(-\lambda/M^2) } \q .
\ee
In the limit $\lambda \rightarrow 0$ most of the $\rho_{N0}$ vanish:
\be
\lim_{\lambda\rightarrow 0}\rho_{N0}(\lambda)= \begin{cases} 2&  \text{for $|N|=1$},\\
                                           0& \text{ for $|N|>1$}.
\end{cases}
\ee
Hence the contribution from the absolutely continuous part of the spectrum 
to 
the physical Hilbert space consists of just two vectors. This contribution 
corresponds to the fact that the delta function $\delta(x_1)\delta(x_2)$ is a generalized eigenvector of $\MCO$.

The total physical Hilbert space is the sum of the contributions from 
$\ch_{pp}$ and $\ch_{ac}$, i.e the sum of $\ch_{pp}$ which can be 
identified with $L_2(\Rl_+,rdr)$ and two vectors $\{e_{10},e_{-10}\}$.
Notice that if we had made the wrong identification of $L_2(\Rl^2,d^2x)$
with $L_2(\Rl_+,rdr)\otimes L_2(S^1,d\phi)$ discussed above then 
we would have had to use the vectors $\Omega'_{N0}\propto e^{in\phi}
e^{-r^2},\;n\not=0$ to get a cyclic system. The spectral measures of these 
vectors all coincide and therefore the corresponding Radon -- Nikodym 
derivatives would all be non vanishing at $\lambda=0$ and the contribution 
to the physcal Hilbert space from the continuous spectrum would be 
infinite dimensional which would be physically wrong because the 
constraint $x_1^2+x_2^2=0$ corresponds to only one point in phase space 
and the correponding physical Hilbert space should be finite dimensional.

\section{Conclusions}
\label{s9}

The main lesson learnt in this article is that the Master Constraint 
Programme can deal essentially with all situations that one usually 
encounters when quantizing finite dimensional systems. It is even possible 
to deal with second class constraints as already emphasized by Klauder
in his Affine Quantization Programme for gravity \cite{2.1}. 
A situation that we have not dealt with are constraints which are 
quadratic in the momenta with structure constants but such that the 
corresponding Lie group is not compact. We will deal with this difficult case 
in a seperate paper \cite{III}. 

In all the examples studied we recover the usual results. This might seem 
surprising because one would think, e.g. that the space of solutions to 
the single quadratic constraint $\MCO=\hat{p}_1^2+\hat{p}_2^2=0$ is larger 
than the space of solutions to the individual linear constraints 
$\hat{p}_1=\hat{p}_2=0$. Indeed, the general solution of the former
is of the form $f_1(z)+f_2(\bar{z})$ where $z=x_1+ix_2$ and $f_1,f_2$ 
are smooth functions while the general solution to the latter are the 
constants where we have used the representation 
$\hat{p}_j=i\hbar\partial/\partial x_j$ on ${\cal H}=L_2(\Rl^2,d^2x)$. 
However, solutions of the first type do not appear in the 
spectral resolution of the Master Constraint. Intuitively, this comes 
about because a physical Hilbert space based on square integrable linear 
combinations of 
holomorphic or antiholomorphic functions must be of the form
$L_2(\Cl,dz d\bar{z} \rho(|z|))$ with a damping factor $\rho(|z|)$.
However, this Hilbert space is not a representation space for a self 
adjoint representation of the Dirac obsrvables with respect to $\MCO$:
Indeed, the induced action of the Dirac observables $\hat{p}_j$ is not 
self-adjoint in this representation. Thus, this representation is not 
viable unless we restrict ourselves to the constant functions. These 
are automatically selected by the spectral analysis of the Master 
Constraint which in turn is induced by the spectral analysis of the 
individual constraints $\hat{C}_j=\hat{p}_j$.

Another point worth mentioning is the following: Assume a simple 
situation, a finite number of first class constraints $C_j$ which may 
close with structure functions only. 
In the case of non -- 
trivial structure functions, the constraints $\hat{C}_j$ {\it 
generically must not 
be self -- adjoint} for the following reason: Let 
$\{C_j,C_k\}=f_{jk}\;^l C_l$ with non -- trivial, real valued structure
functions $f_{jk}\;^l$. Suppose that the $\hat{C}_j$ are self --adjoint
then in quantum theory we expect a relation of the form
$[\hat{C}_j,\hat{C}_k]=i\hbar(\hat{f}_{jk}\;^l \hat{C}_l+\hat{C}_l 
(\hat{f}_{jk}\;^l)^\dagger)/2$ where the symmetric ordering is forced on 
us 
due to the antisymmetry of the commutator. Since $f_{jk}\;^l$ is real 
valued $\hat{f}_{jk}\;^l-(\hat{f}_{jk}\;^l)^\dagger$ should vanish
as $\hbar\to0$. Now 
suppose that $\Psi$ is a 
generalized solution\footnote{More precisely we should look for solutions 
in a space of distributions $\Phi^\ast$ dual to a dense and invariant 
(under the constraints) subspace $\Phi\subset {\cal H}$ in the form 
$\Psi(\hat{C}_j^\dagger f)=0$ for all $f\in \Phi$ and all $j$.}
of all constraints, $\hat{C}_j \Psi=0$ for all $j$. Applying the 
commutator we find $0=i\hbar[\hat{C}_l,(\hat{f}_{jk}^l)^\dagger]\Psi$ for 
all $j,k$. Now as $\hbar\to 0$, the expression 
$i\hbar[\hat{C}_l,(\hat{f}_{jk}^l)^\dagger]$ becomes the Poisson bracket
$\hbar^2\{C_l,f_{jk}\;^l\}$ to lowest order in $\hbar$. Unless this 
classical quantity vanishes, the solution $\Psi$ not only satisfies 
$\hat{C}_j\Psi=0$ but the additional constraints 
$[\hat{C}_l,(\hat{f}_{jk}^l)^\dagger]\Psi=0$ which have no classical 
counterpart. Iterating like this it could happen, in the worst case, that
$\Psi$ satisfies an indefinite tower of new constraints, leaving us 
with the only solution $\Psi=0$. In general the solution space will be too 
small as to capture the physics of the classical reduced phase space, 
displying a quantum anomaly.
In the example with structure functions studied in this paper it was 
actually the case that 
$\{C_l,f_{jk}\;^l\}=0$ which is why we succeeded in using self adjoint 
constraint operators there\footnote{
We should study another example in the future where this is not 
the case.}. 
%A simple example would be $C_j=\epsilon_{jkl} x_k p_l,\;j=1,2$
%These constraints close with (unfortunately singular) structure functions 
%$\{C_1,C_2\}=C_3=-(x_1 C_1+x_2 C_2)/x_3$, however, here it is again 
% satisfied. 
This argument just given is not new, it was 
presented first in \cite{HK}, but it is often forgotten. 

Finally, let us assume that zero lies in the point spectrum 
of all the $\hat{C}_j$ for a given Hilbert space representation and 
set $\MCO:=\sum_j \hat{C}_j^\dagger \hat{C}_j$. Now the eigenvalue 
equations $\hat{C}_j\Psi=0$ and $\MCO\Psi=0$ make sense for elements 
$\Psi\in {\cal H}$. Then we claim that that $\hat{C}_j\Psi=0$ for all $j$
if and only if $\MCO\Psi=0$. The first implication is clear. For the 
converse we compute $0=<\Psi,\MCO\Psi>=\sum_j ||\hat{C}_j\Psi||^2$.
Thus we see that in this case the Master Constraint Operator captures 
exactly the same physics as the individual constraints, in particular it 
will suffer under potentially the same quantum anomalies. Hence, nothing 
is swept under the rug\footnote{However, it is interesting to notice that 
by subtracting the minimum of the spectrum from the Master Constraint
we may also be able to deal with anomalous situations!}.
%This could be 
%tested for instance by applying the Master Constraint to string theory
%using the usual anomalous Fock representation for the Virasoro 
%constraints.}. 
In case that zero also lies in the continuous 
spectrum, the DID procedure of \cite{I} generalizes the argument just 
made.\\ 
\\
\\
\\
{\large Acknowledgements}\\
\\
BD thanks the German National Merit Foundation for financial support.
This research project was supported in part by 
a grant from NSERC of Canada to the Perimeter Institute for Theoretical 
Physics.


\begin{thebibliography}{99}

\parskip -5pt

\bibitem{7.0} T. Thiemann, ``The Phoenix Project: Master Constraint 
Programme for Loop Quantum Gravity'', gr-qc/0305080

\bibitem{I} B. Dittrich, T. Thiemann, ``Testing the Master Constraint 
Programme for Loop Quantum Gravity I. General Framework'',
gr-qc/0411138

\bibitem{III} B. Dittrich, T. Thiemann, ``Testing the Master Constraint 
Programme for Loop Quantum Gravity III. SL(2,R) Models'',
gr-qc/0411140

\bibitem{IV} B. Dittrich, T. Thiemann, ``Testing the Master Constraint 
Programme for Loop Quantum Gravity IV. Free Field Theories'',
gr-qc/0411141

\bibitem{V} B. Dittrich, T. Thiemann, ``Testing the Master Constraint 
Programme for Loop Quantum Gravity V. Interacting Field Theories'',
gr-qc/0411142

\bibitem{1.1} 
C. Rovelli, ``Loop Quantum Gravity", Living Rev. Rel. {\bf 1} (1998) 1,
gr-qc/9710008\\
T. Thiemann,``Lectures on Loop Quantum Gravity'', Lecture Notes in 
Physics, {\bf 631} (2003) 41 -- 135, gr-qc/0210094\\
A. Ashtekar, J. Lewandowski, ``Background Independent Quantum Gravity:
A Status Report'', Class. Quant. Grav. {\bf 21} (2004) R53; 
[gr-qc/0404018]\\
L. Smolin, ``An Invitation to Loop Quantum Gravity'', hep-th/0408048

\bibitem{7.2} C. Rovelli, ``Quantum Gravity'', Cambridge University Press,
Cambridge, 2004

\bibitem{7.3} T. Thiemann, ``Modern Canonical Quantum General 
Relativity'', Cambridge University Press, Cambridge, 2005,
gr-qc/0110034

\bibitem{7.1} T. Thiemann, ``Anomaly-free Formulation of non-perturbative,
four-dimensional Lorentzian Quantum Gravity", Physics Letters {\bf B380}
(1996) 257-264, [gr-qc/9606088]\\
T. Thiemann, ``Quantum Spin Dynamics (QSD)",
Class. Quantum Grav. {\bf 15} (1998) 839-73, [gr-qc/9606089];
``II. The Kernel of the Wheeler-DeWitt Constraint Operator",
Class. Quantum Grav. {\bf 15} (1998) 875-905, [gr-qc/9606090];
``III.
Quantum Constraint Algebra and Physical Scalar Product in Quantum General
Relativity", Class. Quantum Grav. {\bf 15} (1998) 1207-1247,
[gr-qc/9705017];
``IV. 2+1 Euclidean Quantum Gravity as a model to test 3+1
Lorentzian Quantum Gravity", Class. Quantum Grav. {\bf 15} (1998) 
1249-1280, [gr-qc/9705018]; 
``V. Quantum Gravity as the Natural Regulator of the Hamiltonian 
Constraint
of Matter Quantum Field Theories",
Class. Quantum Grav. {\bf 15} (1998) 1281-1314, [gr-qc/9705019];
``VI. Quantum Poincar\'e Algebra and a Quantum Positivity of Energy
Theorem for Canonical Quantum Gravity",
Class. Quantum Grav. {\bf 15} (1998) 1463-1485, [gr-qc/9705020];
``Kinematical Hilbert Spaces for Fermionic and
Higgs Quantum Field Theories",
Class. Quantum Grav. {\bf 15} (1998) 1487-1512, [gr-qc/9705021]


\bibitem{8.2} 
A. Ashtekar, J. Lewandowski, D. Marolf, J. Mour\~ao, T.
Thiemann, ``Quantization for diffeomorphism invariant theories
of connections with local degrees of freedom", Journ. Math. Phys.
{\bf 36} (1995) 6456-6493, [gr-qc/9504018]

\bibitem{7.4} H. Sahlmann, ``When do Measures on the Space of Connections
Support the Triad Operators of Loop Quantum Gravity?'', gr-qc/0207112;
``Some Comments on the Representation Theory of the Algebra Underlying
Loop Quantum Gravity'', gr-qc/0207111\\
H. Sahlmann, T. Thiemann, ``On the Superselection Theory of
the Weyl Algebra for Diffeomorphism Invariant Quantum Gauge Theories'',
gr-qc/0302090;
``Irreducibility of the Ashtekar-Isham-Lewandowski Representation'',
gr-qc/0303074\\
A. Okolow, J. Lewandowski, ``Diffeomorphism Covariant
Representations of the Holonomy Flux Algebra'', gr-qc/0302059





\bibitem{2.1} J. Klauder, ``Universal Procedure for Enforcing Quantum 
Constraints'', Nucl.Phys.B547:397-412,1999, [hep-th/9901010];
``Quantization of Constrined Systems'', Lect. Notes Phys. 
{\bf 572} (2001) 143-182, [hep-th/0003297]\\
A. Kempf, J. Klauder, ``On the Implementation of Constraints through 
Projection Operators'', J. Phys. {\bf A34} (2001) 1019-1036,
[quant-ph/0009072]


%\bibitem{2.2}  I. M. Gel'fand, N. Ya. Vilenkin, ``Generalized Functions",
%            vol. 4, Applications of Harmonic Analysis, Academic Press,
%            New York, London, 1964

\bibitem{6.1} J.E. Marsden, P.R. Chernoff, ``Properties of Infinite
Dimensional Hamiltonian Systems", Lecture Notes in Mathematics,
Springer-Verlag, Berlin, 1974

\bibitem{8.1} T. Thiemann, ``Quantum Spin Dynamics (QSD): VIII.
The Master Constraint'', in preparation


\bibitem{8.3} H. Sahlmann, T. Thiemann, ``Towards the QFT on
Curved Spacetime Limit of QGR. 1. A General Scheme'', [gr-qc/0207030];
``2. A Concrete Implementation", [gr-qc/0207031]

%\bibitem{8.4} E. Witten, Nucl. Phys. {\bf B311} (1988) 46

\bibitem{8.5} T. Thiemann, ``Quantum Spin Dynamics (QSD): VII.
Symplectic Structures and Continuum Lattice Formulations of
Gauge Field Theories", Class.Quant.Grav.18:3293-3338,2001,
[hep-th/0005232]; ``Gauge Field Theory Coherent States (GCS): I.
General Properties", Class.Quant.Grav.18:2025-2064,2001, [hep-th/0005233];
``Complexifier Coherent States for Canonical Quantum General Relativity", 
gr-qc/0206037\\
T. Thiemann, O. Winkler, ``Gauge Field Theory Coherent States
(GCS): II. Peakedness Properties", Class.Quant.Grav.18:2561-2636,2001,
[hep-th/0005237]; ``III. Ehrenfest Theorems",
Class. Quantum Grav. {\bf 18} (2001) 4629-4681, [hep-th/0005234];
``IV. Infinite Tensor Product and Thermodynamic Limit",
Class. Quantum Grav. {\bf 18} (2001) 4997-5033, [hep-th/0005235]\\
H. Sahlmann, T. Thiemann, O. Winkler, ``Coherent States for
Canonical Quantum General Relativity and the Infinite Tensor Product
Extension", Nucl.Phys.B606:401-440,2001
[gr-qc/0102038]

\bibitem{RS} M. Reed, B. Simon, ``Methods of Modern Mathematical Physics I: Functional Analysis'', Academic 
Press, New York, 1980

\bibitem{PartObs1} B. Dittrich, ``Partial and Complete Observables for Hamiltonian Constrained Systems'', gr-qc/0411013


%\bibitem{ReedSimonI}Reed,Simon I
\bibitem{ReedSimonII}M. Reed, B. Simon, ``Methods of Modern Mathematical Physics II: Fourier Analysis, Self-Adjointness'', Academic 
Press, New York, 1975

\bibitem{SchroedOp} G. Teschl, ``Schr\"odinger Operators'', lecture notes for a cours at the University Vienna 1999 and 2002, available at {http://www.mat.univie.ac.at/{${}^\sim$}gerald/ftp/book{${}^\sim$}schroe/}

\bibitem{howe} R. Howe,E. C. Tan, ``Non-Abelian Harmonic Analysis -- Applications of SL(2,R)'', Springer 1992, New-York

\bibitem{Galindo} A. Galindo, P. Pascual, ``Quantum Mechanics I'', (Springer, Berlin) 1990

\bibitem{CH} R. Courant, D. Hilbert, ``Methods of Mathematical Physics'', 
vol. 1, Interscience Publishers, New York, 1953

\bibitem{HK} P. Haj\'i\v{c}ek, K. Kucha\v{r}, Phys. Rev. {\bf D41}
(1990) 1091, Journ. Math. Phys. {\bf 31} (1990) 1723
% symm versus non-symm Ham. const.



\end{thebibliography}
\end{document}